# Blood ties: ABO is a trans-species polymorphism in primates

Laure Ségurel[1,2,*,#], Emma E. Thompson[1,*], Timothée Flutre[1,3], Jessica Lovstad[1], Aarti Venkat[1], Susan W. Margulis[4,a], Jill Moyse[4], Steve Ross[4], Kathryn Gamble[4], Guy Sella[5], Carole Ober[1,+,#], Molly Przeworski[1,2,6,+,#]

[1] Department of Human Genetics, University of Chicago, Chicago, IL, 60637, USA

[2] Howard Hughes Medical Institute, University of Chicago, Chicago, IL, 60637, USA

[3] Department of Genetics and Plant Breeding, INRA, UR1164 URGI, Versailles, 78026, France

[4] Lincoln Park Zoo, Chicago, IL, 60614, USA

[5] Department of Ecology, Evolution and Behavior, The Alexander Silberman Institute of Life Sciences, The Hebrew University of Jerusalem, Jerusalem, 91904, Israel

[6] Department of Ecology and Evolution, University of Chicago, Chicago, IL, 60637, USA

[*] *These authors contributed equally*

[+] *These authors co-supervised the project*

[#] *To whom correspondence should be addressed*

[a] *Current address: Canisius College, Buffalo, NY, USA*

**Corresponding authors:**

Laure Ségurel
Email: lsegurel@uchicago.edu
Carole Ober
Email: c-ober@bsd.uchicago.edu
Molly Przeworski
Email: mfp@uchicago.edu

**Abstract**

The ABO histo-blood group, the critical determinant of transfusion incompatibility, was the first genetic polymorphism discovered in humans. Remarkably, ABO antigens are also polymorphic in many other primates, with the same two amino acid changes responsible for A and B specificity in all species sequenced to date. Whether this recurrence of A and B antigens is the result of an ancient polymorphism maintained across species or due to numerous, more recent instances of convergent evolution has been debated for decades, with a current consensus in support of convergent evolution. We show instead that genetic variation data in humans and gibbons as well as in Old World Monkeys are inconsistent with a model of convergent evolution and support the hypothesis of an ancient, multi-allelic polymorphism of which some alleles are shared by descent among species. These results demonstrate that the ABO polymorphism is a trans-species polymorphism among distantly related species and has remained under balancing selection for tens of millions of years – to date, the only such example in Hominoids and Old World Monkeys outside of the Major Histocompatibility Complex.

**Background**

Balancing selection pressures can maintain two or more alleles in the population for long periods of time – so long that the polymorphism may be shared identical by descent among distinct species, leading to a trans-species polymorphism. In this scenario, the time to the most recent common ancestor (tMRCA) of the selected alleles will predate speciation times. Due to linkage, sites near the selected ones may also have old tMRCAs, resulting in unusually high diversity within species and shared alleles between species (1). Because of recombination events between the two allelic classes in the history of the sample, however, the tMRCA at linked sites can also be much more recent than at the selected sites, so diversity levels need not be unusually high. Moreover, the old tMRCA provides many opportunities for recombination, which will erode the segment that carries the high diversity signal (2, 3); as a result, ancient balancing selection will leave only a narrow footprint in genetic variation data, and will often be hard to detect (1, 3).

Perhaps for this reason, only a handful of examples of trans-species polymorphisms are known at the molecular level. The most famous examples are the self-incompatibility loci in plants and the Major Histocompatibility Complex (MHC) in vertebrates. The self-incompatibility loci prevent self-fertilization, and variation is likely maintained by negative frequency-dependent selection (4, 5). In turn, the nature of selective pressures at MHC are unclear, but are believed to result from its central role in the recognition of pathogens (for a recent review, see 6). Within humans, MHC is the only region known to harbor variation shared identical by descent with other species. Variants at other loci have also been found to be shared with Hominoid species, potentially indicating shared balancing selection pressures, but patterns of genetic variation do not support the hypothesis that the alleles are identical by descent as opposed to due to recurrent mutations (e.g., the PTC alleles in humans and chimpanzees 7).

Arguably the best studied case of apparent convergent evolution is the ABO blood group (1, 8), the first molecular polymorphism to be characterized in humans. ABO blood groups (A, B, AB, O) are defined by the presence or absence of specific antigens that circulate in body fluids and are attached to lipids at the surface of various epithelial and endothelial cell types (notably in the gastro-intestinal tract, but also, in Hominoids only, on red blood cells) (9, 10). These antigens are associated with complementary immune antibodies produced in the gut after contact with bacteria and viruses carrying A-like and B-like antigens (11). While the biological significance of ABO outside of its role in transfusion is unclear (12), histo-blood antigens in general are known to act as cellular receptors by which pathogens can initiate infections (13-15). Furthermore, variation in ABO has been associated with susceptibility to a number of infectious diseases (reviewed in ref. 16), pointing to a role of ABO in immune response.

The presence or absence of the A, B and H (O) antigens result from allelic variation at the *ABO* gene, which encodes a glycosyltransferase. The A transferase, able to transfer N-acetyl-D-galactosamine to the H acceptor substrate, is encoded by the A allele and the B transferase, able to transfer D-galactose to the same acceptor substrate, by the B allele. A and B alleles at *ABO* are co-dominant whereas the null O alleles are recessive (17). Two amino acids at positions 266 and 268 in exon 7 are responsible for the A and B enzymatic specificity in humans (18), and are surrounded by a peak of nucleotide diversity (3, 19, 20). As an illustration, the diversity level in exon 7 is 0.0058 in a Yoruba sample, a value equaled or exceeded in only 0.08% of comparable exonic windows in the genome (see Methods). This peak of diversity provides support for long-lived balancing selection acting on this locus (20, 21).

Remarkably, the A, B and H antigens exist not only in humans but in many other primates (reviewed in 22), and the same two amino acids are responsible for A and B enzymatic specificity in all sequenced species (8, 18, 23-25). Thus, primates not only share their ABO blood group, but also the same genetic basis for the A/B polymorphism. O alleles, in contrast, result from loss of function alleles such as frame-shift mutations, and appear to be species-specific (26). That different species share the same two A/B alleles could be the result of convergent evolution in many lineages or of an ancestral polymorphism stably maintained for millions of years and inherited across (at least a subset of) species. The two possibilities have been debated for decades, with a consensus emerging that A is ancestral and B allele has evolved independently at least six times in primates (in human, gorilla, orangutan, gibbon/siamang, macaque and baboon) (8, 25, 26); in particular, that the human A/B polymorphism arose more recently than the split with chimpanzee (8, 20) (see SOM). We show that instead the remarkable distribution of *ABO* alleles across species reflects the persistence of an old ancestral polymorphism that originated at least 20 Millions of years (My) ago and is shared identical by descent by humans and gibbons as well as among distantly related Old World Monkeys.

**Results**

Previous to ours, 31 studies reported phenotypic and/or genetic data on ABO in non-human primates (Supplementary Table 1). To supplement these data, we sequenced exon 7 of the *ABO* gene in four Hominoid species (10 bonobos, 35 western chimpanzees, 31 lowland gorillas, and 12 orangutans from both Sumatra and Borneo), in two previously uncharacterized Old World Monkeys (five colobus and four vervet monkeys), as well as, for the first time, in two New World Monkeys (four marmosets and three black howler monkeys) (see Methods, SOM and Supplementary Table 2 and 3). Among 40 non-human species, the ABO polymorphism is ubiquitous, with 19 species in 10 genera polymorphic for A and B, as well as 10 species in seven genera without a B in our sample and 11 species in six genera without an A (Supplementary Table 1 and Figure 1); 15 species in 11 genera also have an unambiguous O allele. Contrasting sequences obtained for marmosets (which are all A in our sample) and black howler monkeys (all B in our sample) reveals that the genetic basis for A/B specificity is the same in New World Monkeys as in Hominoids (see Figure 1). For the A allele, we further observed that, compared to Hominoids, colobus and vervet monkeys use a different codon to encode the same amino acid at one of the two functional positions, as was previously noted for macaques (24) and baboons (27) (see Figure 1). This phylogenetic pattern is consistent with a synonymous substitution in the A lineage leading to Old World Monkeys.

To distinguish between the maintenance of an old trans-species polymorphism (Figure 2a) and more recent convergent evolution (Figure 2b), we first sought to gain a sense of the length of the segment that should carry a signal of a trans-species polymorphism. To this end, we approximated the expected length of the (two-sided) segment contiguous to the focal sites, for plausible values of the salient parameters (see Methods for details). Consistent with previous modeling work (2), we found that it should be at most a few hundred base pairs (bps) in length, depending notably on the pair of species considered and the recombination rate (Figure 2c; Supplementary Figure 1). One implication is that previous studies that considered larger windows or segments far from the selected sites may have missed the footprints of a trans-species polymorphism (see SOM).

Based on our calculations, we generated trees of *ABO* haplotypes for short regions around the functional sites, i.e., 300 bps in Hominoids and 200 bps in Old World Monkeys. Strikingly, in the trees within Hominoids (other than orangutan) and within Old World Monkeys, A and B alleles cluster by type rather than by species (Figure 3). The clustering by A and B types is consistent with a scenario in which the A and B allelic classes had a most recent common ancestor long ago and persisted across species (Figure 2a). The lack of clustering for the O alleles could then reflect frequent turnovers (i.e., replacement) of these alleles within a balanced polymorphism, as expected from the high mutation rate to null alleles (28). Alternatively, the phylogenetic trees could reflect convergent evolution of multiple functional mutations in numerous lineages (Figure 2b).

Because of recombination, diversity among *ABO* alleles should be highly heterogeneous along the sequence, with at most a few (and possibly no) short segments showing unusually high diversity levels (2; see Methods). We therefore focused on small, sliding windows along exon 7 (see Supplementary Figure 2-3 for a slightly larger window choice), comparing the pairwise synonymous diversity among *ABO* allelic classes to the 95% confidence interval for synonymous divergence between the same allele (A or B) from different species (see Methods). In comparisons of lineages from different species, a model of convergent evolution predicts that, near the selected sites, the divergence between allelic classes (e.g., A and B) will be the same as divergence within allelic classes (e.g., A and A), whereas a model of trans-species polymorphism predicts that the former will exceed the latter. The excess will be subtle, however, because the divergence between lineages from the same allelic class sampled in different species (e.g., A and A) is also inflated close to an ancient balanced polymorphism (due to recombination; see Figure 1 in ref. 2). We note further that because O is a null allele, there is a high mutation rate to O, such that the current O could be derived from any functional allele, including some that have not persisted to the present. As a consequence, divergence between O and A (or O and B) may be deep, possibly even deeper than A versus B, even if the O allele is itself recent.

Within Old World Monkeys, there is tremendous synonymous diversity between A and B alleles in macaques, exceeding the divergence between macaque and baboon, and consistent with the divergence between macaque and colobus monkey ~18 Mya (29) (Figure 4a). Similarly high synonymous diversity is visible between A and B alleles within colobus monkeys, notably around the functional sites (Figure 4a). A similar but weaker pattern is seen in baboons (Figure 4a), possibly due to greater erosion of the ancestral segment by recombination. In accordance with the inheritance of an ancestral segment identical by descent across macaques and baboons, two synonymous polymorphisms are shared between the two species (Figure 4a). Further evidence that the origin of the polymorphism predates their split comes from the comparison between baboon A and macaque O (or baboon O and macaque B), which reveal deeper coalescent times between allelic classes than seen within an allelic class (i.e., baboon and macaque A or B). Together, these findings strongly support the hypothesis of a trans-species polymorphism shared among macaque, baboon and colobus monkey, with an origin as old as 18 Mya (or greater).

In Hominoids, in turn, human *ABO* diversity is over an order of magnitude higher than typical polymorphism levels of ~0.1% (30) (Figure 4b), with synonymous diversity between *ABO* alleles similar to divergence levels among African apes and compatible with a divergence as high as between human and gibbon lineages, who last had a common ancestor ~20 Mya (29). This situation is mirrored in gibbons (Figure 4b), as expected if humans and gibbons share A and B alleles identical by descent. Furthermore, divergence between chimpanzee A and human B alleles significantly exceeds the divergence among African ape lineages (Figure 4b), which is predicted only if the A/B polymorphism predates the species' split. In orangutan, in contrast, A and B lineages are highly similar (Supplementary Figure 3), suggesting they are recently derived; since this case could represent a turnover event in this lineage, however, it does not preclude the possibility that the A/B balanced polymorphism is older (28). Thus, synonymous pairwise differences in Hominoids point to an origin of the A/B polymorphism before the divergence of extant African apes ~8 Mya, and are consistent with the maintenance of a balanced polymorphism since the root of Hominoids ~20 Mya, which persisted in humans and gibbons.

Because of the stochasticity of recombination and mutation events in a small window, the divergence levels are noisy and do not establish whether the polymorphism is significantly older than 8 My old, i.e., whether A and B alleles arose twice in Hominoids, once in the ancestor of the African apes and once in the ancestor of gibbons and siamangs, or only once, before the human-gibbon split. We therefore considered an additional prediction that allows us to distinguish between these two cases. When multiple allelic classes are maintained in the population, new mutations can drift up to high frequency within one class, but cannot fix in the species until they recombine onto the other class (31). If A and B date back to the origin of Hominoids ~20 Mya, then the maintenance of the two allelic classes will have reduced the number of synonymous fixations at linked sites relative to what would be expected in the absence of a balanced polymorphism; in other words, the deeper coalescent times at linked sites will have left less time for fixations to occur. In contrast, under a model in which B (or A) evolved twice, only one allelic class would have been present in the first 12 My of Hominoid evolution (whether A or B), and no such slowdown in synonymous fixations would be expected. We tested the null hypothesis that only one *ABO* allelic class was present during the first 12 My of Hominoid evolution by comparing the observed rate of synonymous substitutions in extant "monomorphic" branches that do not have both A and B (i.e., chimpanzee, gorilla and siamang) to the rate in the two internal branches of the Hominoid phylogeny (using 585 bps of exon 7; Figure 4c, see Methods). We found that the internal branches evolved significantly more slowly than did monomorphic lineages ($p$-value = 0.01). In contrast, we could not reject a null model in which the two internal branches evolved at the same rate as did branches polymorphic for A and B (human and gibbon; $p$-value = 0.18). Thus, the data are inconsistent with a model where only one allelic class existed during the first 12 My of Hominoid evolution (i.e., convergent evolution in humans and gibbons) and are best explained by the persistence of the A and B allelic class for the ~20 My since divergence of humans and gibbons.

**Discussion**

Our results indicate that ABO is a trans-species polymorphism inherited identical by descent in humans and gibbons as well as among all Old World Monkeys studied (macaques, baboons and colobus monkeys), and which was therefore maintained over tens of millions of years. ABO is the second example of a locus at which human variation traces back to the origin of Hominoids. Moreover, given that the signal of a trans-species polymorphism is expected to decrease over time, eventually becoming undetectable (2, 3) (Figure 2c), we cannot exclude an older origin of the balanced polymorphism, which led to allele sharing among Hominoids and Old World Monkeys (albeit with a turnover of the A allele), or possibly even with New World Monkeys. This finding points to selection pressures that have remained strong relative to genetic drift throughout the evolution of these species (28).

As remains the case for the MHC (6), the selection mechanism maintaining this polymorphism across so many primate species is largely unknown. One possibility, heterozygote advantage (seen e.g., in the *HBB* gene in response to malaria and sickle-cell anemia 32), may be unlikely to underlie long-lived balancing selection, instead representing a transient solution to balancing selection pressures until a single allele that confers the heterozygote phenotype arises or a duplication occurs (33). In support of this argument, the AB phenotype can be created by single "cis-AB" alleles (34) and yet such alleles have not reached fixation in any of the surveyed species and are rare in humans (see Supplementary Table 2). Moreover, the presence of the ABO polymorphism throughout primates implies that the selected phenotype is probably not tied to the expression of antigens on red blood cells, a trait restricted to Hominoids (35). Humans are known to have many histo-blood subgroups (notably among A types), which are interchangeable for transfusion purposes, but differ in quantity and quality of antigens (36, 37); similarly, chimpanzees, gorillas, orangutans and gibbons have been reported to have variable A and B subgroups, respectively, that differ in antigenic properties (26, 38). Thus, although A, B, and O are clearly of functional importance and may denote the strongest fitness differences among variants of the *ABO* gene, these histo-blood labels are unlikely to provide a complete description of the allelic classes acted on by natural selection. These considerations suggest that variation at *ABO* reflects a multi-allelic balanced polymorphism, with cryptic differences in function among A and among B alleles.

As expected from an ancient multi-allelic balanced polymorphism (21, 28), there was occasional turnover within allelic classes, including of a codon in the lineage leading to OWM and of A and B alleles within orangutan, as well as frequent turnovers of O alleles in all species (28). Numerous losses of A and B have also occurred (see Figure 1), possibly as a result of bottlenecks in the history of the species (39) or due to differences in selection pressures among lineages.

In any case, the maintenance of ABO across so many primates reveals previously unknown and important functions of a heavily studied gene. More generally, this study illustrates a general approach that can be used to scan for ancient balancing selection in the genome and raises the possibility that, with the availability of genome-wide polymorphism data from closely related species, this mode of selection will turn out to be more common than currently believed.

**Methods**

*Genome-wide estimate of exonic diversity in 585 bps sliding windows.* We used the primate orthologous exon database (http://giladlab.uchicago.edu/orthoExon/) to obtain the coordinates of unique, non-overlapping orthologous exons between human and rhesus macaque genomes. This dataset represents about 29% of human exons and does not include the MHC region. We estimated human diversity in the YRI sample (individuals of Yoruban ancestry) from the 1000 Genome Pilot project (30). For each exon, we calculated the mean diversity in humans in a 585 bps sliding window. Windows with a gap greater than 10% of the sequence were excluded. We then considered the fraction of windows with diversity levels higher than in *ABO* exon 7 (where the diversity was 0.0058) and obtained a value of 0.08%. To control for differences in mutation rate, we verified that a similarly low *p*-value was obtained for the ratio of diversity to divergence between human and rhesus macaque (this ratio is 0.16 in *ABO* exon 7; *p*-value = 0.2%). The *p*-values obtained in this way are likely to be over-estimates, as some of the more extreme windows are probably the result of errors in SNP calls or in read mapping.

*Re-sequencing data for ABO.* We used previously published data: in humans, 60 CEU individuals of European ancestry, 60 CHB + JPT individuals of South-East Asian ancestry and 59 YRI individuals of Sub-Saharan African ancestry (30), 31 olive baboons (*Papio anubis*) (27), 13 macaques (seven cynomolgus, or crab-eating, macaques, *Macaca fascicularis,* and six rhesus macaques, *Macaca mulatta*) (24), 17 gibbons (five agile gibbons, *Hylobates agilis*, 12 white-handed gibbons, *Hylobates lar*) and six siamangs (*Symphalangus*

*syndactylus*) (25).

In addition, we sequenced Hominoids samples from Lincoln Park Zoo (40) (see SOM): 10 bonobos (*Pan paniscus)*, 35 western chimpanzees (*Pan troglodytes)*, 31 lowland western gorillas (*Gorilla gorilla*) and nine orangutans (*Pongo pygmaeus*: three orangutans from Sumatra, two from Borneo and four hybrids). Three black howler monkeys (*Alouatta caraya*) samples were also obtained from Lincoln Park Zoo. Additionally, samples from three Sumatran orangutans were purchased from the San Diego Zoo, five colobus monkeys (three *Colobus angolensis*, one *Colobus polykomos* and one *Colobus guereza*) from the IPBIR (Integrated Primate Biomaterials and Information Resource) through the Coriell Institute, four vervet monkeys (*Chlorocebus aethiops*) from Alpha Genesis, and four marmosets (*Callithrix jacchus*) from the Southwest Foundation for Biomedical Research.

When needed, genomic DNA was extracted from blood using the Puregene DNA Isolation Kit (Gentra Systems). Amplification of exon 7 was performed using primers and conditions described in the Supplementary Table 4. Sequencing reactions were performed using the Big Dye Terminator v3.1 Cycle Sequencing Kit (Applied Biosystems). Chromatograms were aligned and analyzed using the Phred-Phrap-Consed package (41).

Haplotypes were estimated in each species separately using PHASE2.1 (42). The program was run twice, with different seeds (option –S); the two outputs were identical other than for six SNPs in colobus monkeys (of which five SNPs were in perfect LD) and one SNP in humans (in the CHB + JPT population). Trees and diversity plots rely on the first output, but identical conclusions were obtained with the second one.

The substitutions and polymorphisms requiring more than one mutation given the accepted species tree (29) are listed in Supplementary Table 2 and the numbers of polymorphic sites found in each species in Supplementary Table 3.

*Identification of the A, B and O haplotypes*. In humans, we used re-sequencing data from exon 6 and exon 7 to classify haplotypes in three main lineages: A, B, or O. Considering the negative strand on chromosome 9 in build hg19, A haplotypes are defined as those carrying the A-specific alleles (codons) at the two functional sites: C at position 136.131.32 (Leucine at amino acid 266) and G at position 136.131.315 (Glycine at amino acid 268). B haplotypes are defined as those carrying an A (Methionine) and a C (Alanine) at these positions. O haplotypes are defined as those carrying the 1 bp deletion between position 136.132.908 and 136.132.907, and almost always carry the A-specific alleles at the two functional sites.

We found two recombinant haplotypes between the two functional sites (at frequency 0.6% in the combined human population samples) and excluded them from subsequent analyses.

In other species, A and B haplotypes were aligned to humans using the Phred-Phrap-Consed package (41) and analogously defined by their amino acids at position 266 and 268. O haplotypes were considered only when defined based on their presence in individuals with an O phenotype or if frame-shift mutations had been clearly identified. In chimpanzee, there are two such O haplotypes (43): Odel, which carries a 9 bp deletion in exon 7 and Ox, which carries a G at nucleotide position 791. An unambiguous O haplotype has been found in macaques and in baboons as well, even though the causal mutation has not been identified (24, 27). Finally, in gibbons, a 7-bp deletion found in exon 7 was considered to define a null allele (25). In baboons (27), an O haplotype (< 5%) has been predicted by comparing the expected phenotype and genotypes of individuals; since this result could arise from phenotypic errors or alleles associated with low expression of antigens, however, we excluded this allele from consideration.

*Estimating the length of the segment carrying a signal of a trans-species polymorphism.* The presence of a trans-species polymorphism at one site distorts patterns of genetic variation data at linked sites, but only over a short distance (2). To estimate this distance, we assumed that balancing selection has maintained two alleles, A and B, at a selected site (without turnover) in two species since before the time of their split, $T$ generations ago. Specifically, we assumed that, at the selected site, the time to the most recent common ancestor for two A alleles or two B alleles sampled from different species is less than the coalescent time for an A and a B allele sampled from different species. We then derived expressions for the expected length of the segment contiguous to the selected site on which we expect to see various signals of a trans-species polymorphism. For brevity, we further assumed that salient parameters (recombination rate, generation time, and frequencies of the selected classes) remained the same in the two lineages as well as in the ancestral population.

First, we considered a sample of two haplotypes from a single species, one from each selected class, and estimated the expected length of the contiguous segment that does not coalesce (backward in time) before time $T$ (also estimated by 2). This is the relevant length scale over which a trans-species polymorphism will leave a signal in *intra-species* comparisons (such as those in Figures 4a and 4b). The expected length can be bound from below by considering the segment on one side of the selected site that experienced no recombination in either sample during $T$ generations. The length of this segment follows an exponential distribution $x \sim Exp(2cT)$, where $c$ is the recombination rate per bp, per generation (cf. 2). Thus, the expectation of the length of the two-sided shared segment is $1/cT$.

Assuming that only recombination events with haplotypes from the other selected class reduce the segment size, then $x \sim Exp(cT)$. This follows because, if allele A has frequency $p$ and allele B frequency $q$ ($p+q=1$), then the distribution of lengths of the segment contiguous to allele A that did not experience recombination with the other class is $x_A \sim Exp(cTq)$ and similarly $x_B \sim Exp(cTp)$; therefore, the length of the segment to experience recombination in either background is $x = \min\{x_A, x_B\} \sim Exp(cT(q+p))$. Thus, the expectation of the length of the two-sided segment is $2/cT$. This calculation neglects the possibility of a more complex sequence of recombination events. For example, consider a recombination event with a haplotype from the same selected class at distance K of the selected site; we ignored the possibility that the recombination event is with a haplotype that previously recombined with the other selected class between the selected site and K. It can be shown that, if $T \gg 2N_e$ generations, which is satisfied for the cases considered here, the contribution of these events is negligible. We note further that, throughout, we considered only one chromosome from a given selected class in a given species. This is appropriate because coalescent times within each of the selected classes should be small compared to the species' split times (i.e., because $T \gg 2N_e$).

Next, we considered a sample of two chromosomes, each from a different species. We asked about the expected length of the contiguous segment in which a pair in the same selected class coalesces before a pair from different selected classes. In such a segment, divergence levels for pairs from the same selected class are expected to be less than or equal to the levels between pairs from different selected class. This is therefore the relevant length scale for the *inter-species* comparisons in Figures 4a and 4b. We assumed, without loss of generality, that the pair from different selected classes is an A from species 1 and a B from species 2, and that the pair from the same selected class consists of the same A from species 1 and an A from species 2. By assumption, at the selected site, the A pair coalesces (in the ancestral population) before the A and B pair. This coalescent order would also hold for the segment contiguous to the selected site that experienced no recombination in any of the three lineages before the A pair coalesced. Because recombination events within the same selected class are unlikely to affect the expected order of coalescence events (assuming T >> $2N_e$ and $2N_a$, where $N_a$ is the effective size of the ancestral population), we derived the expected length of the segment by requiring that none of the three lineages recombine with haplotypes from the other selected class. The coalescence time of the A-samples in the ancestral population follows an exponential distribution $t \sim Exp\{1/2N_ap\}$. Conditional on $t$, the length of the one sided segment of interest is exponentially distributed with $x = min\{x_1^A, x_2^A, x_2^B\} \sim Exp\{c(T+t)(2q+p)\}$. The two-sided expected length is therefore

$$E(x) = 2\int_0^\infty E(x|t)f(t)dt = 2\int_0^\infty \frac{1}{c(2-p)(T+t)} \frac{e^{-t/2N_ap}}{2N_ap} dt = \frac{2}{c(2-p)} \int_0^\infty \frac{e^{-z}}{(T+(2N_ap)z)} dz.$$

When $T > 2N_a$, this length increases with $p$ and is therefore bound by the values at $p = 0$ and $p = 1$. In Figure 2c and Supplementary Figure 1, we plot this equation for $p = 0.5$.

Last, we considered a sample of four haplotypes, one from each selected class in each of the two species. We asked about the expected length of the contiguous segment in which the coalescent order of the four haplotypes clusters by selected class (as opposed to by species or neither). This is the relevant scale at which to expect a tree that clusters by selected class and not by species, and the scale on which we would expect to see shared polymorphisms between the two species (besides the selected site). Thus, it is the relevant scale for Figure 3. In this case, we were interested in the time at which both A lineages and both B lineages have coalesced at the selected site, but A and B lineages have not. This time is given by $t = max\{t_A, t_B\}$, where $t_A \sim Exp\{1/2N_ap\}$ and $t_B \sim Exp\{1/2N_aq\}$. The segment (contiguous to the selected site) in which none of the four lineages recombined with the other selected class will have the same tree topology as the selected site, i.e., in it, the haplotypes will group by selected class. Conditional on $t$, the length of the one-sided segment is exponentially distributed with $x = min\{x_1^A, x_2^A, x_1^B, x_2^B\} \sim Exp\{2c(T+t)\}$. In turn, the two-sided expected segment length is

$$E(x) = 2\int_0^\infty E(x|t)f(t)dt = \frac{1}{c}\int_0^\infty \frac{1}{T+t}\left[\left(1-e^{-t/2N_ap}\right)\frac{e^{-t/2N_aq}}{2N_aq} + \left(1-e^{-t/2N_aq}\right)\frac{e^{-t/2N_ap}}{2N_ap}\right]dt.$$

When $T > 2N_a$, this length is maximized for $p = q = 0.5$ and minimized for $p$ or $q = 0$. This equation yields an expectation similar to the one plotted in Figure 2c (see Supplementary Figure 4).

Within this segment, the number of shared SNPs will depend on the age of the balanced polymorphism. On average, there should be more differences between selected backgrounds than expected from the heterozygosity in the ancestral population (since coalescent times between haplotypes from different selected classes are older than average). However, because of recombination in the ancestral population, the contiguous segment becomes shorter and shorter with increasing age of the balanced polymorphism. For a very old polymorphism, the density of shared SNPs will therefore increase with decreasing distance to the selected site.

To estimate the expected segment length for specific species pairs, we used a generation time of 15 and an ancestral effective population size $N_a$ of ~30,000, a typical value for ancestral Hominoids (44, 45). This value is also roughly equivalent to the current effective population size of rhesus macaques (~36.000, based on diversity estimates of 0.29% per bp (46), and a mutation rate of $2x10^{-8}$ per bp per generation, which is higher than in Hominoids and consistent with the higher synonymous divergence in OWM). This value of $N_a$ leads to an expected pairwise coalescence time of $2gN_a$ = 0.9 My, and, using divergence times from ref. (25) and ref. (29), to the following split time estimates: ~7 My for macaque-baboon, ~17 My for macaque-colobus, ~31 My for macaque-human and ~19 My for human-gibbon. In turn, the split between human and chimpanzee was taken to be ~5 My, between human-gorilla ~7 My and between human-orangutan ~12 My (45, 47-50).

*Tree representations of the data.* Based on our expectations for the expected length of the segment in which we expect haplotypes to cluster by selected class (see above), we used a sequence of 200 and 300 bps to generate the trees of Old World Monkeys and Hominoids, respectively. Trees were based on nucleotides 649 to 948 and nucleotide 699 to 898 in Hominoids and Old World Monkeys, respectively, starting at the first ATG codon in the mRNA sequence (Genbank NM_020469.2). In humans, we excluded rare recombinants that carried the deletion in exon 6 together with B-specific alleles at the two functional sites (1.3% in the combined human population samples) as well as haplotypes with recombination events between the A and the B background within the window considered (0.3%).

Phylogenetic trees were obtained with the program MrBayes version 3.1.2 (51). We used a general time reversible model of molecular evolution with a proportion of invariable sites, and a gamma-shaped distribution of rates across sites; other parameters were left at default settings. Using the same rates of substitutions for all nucleotides instead did not affect our results. The program was run three times, each time with two simultaneous Markov chains running for 3,000,000 generations, discarding the initial 25% of the trees as burn-in. The three replicates gave identical consensus trees, with a tight range of credibility values for the clade of B alleles. Trees were plotted with TreeView version 0.5.0 (52).

*Calculating synonymous pairwise differences (dS).* The average synonymous pairwise differences between alleles (dS) per 201 bps (Fig 4a-b) and 300 bps (Supplementary Figure 2 and 3) sliding window in *ABO* exon 7 were calculated with the maximum likelihood method (53) implemented in the program *codeml* from PAML (54). The codon substitution model was the F2x4 model, in which the equilibrium codon frequencies are estimated from the average nucleotide frequencies in the sequence (option CodonFreq=1). The pairwise comparison was used (runmode = -2), to avoid relying on an underlying species tree for all the sequences. To obtain a more precise estimate, the transition to transversion ratio was estimated for all species altogether from 585 bps (the sequence for which we have data for all species). The obtained value (kappa=5.8) was then fixed when estimating the dS per window. Using the approximation from ref. (55) implemented in the program yn00 from PAML (54) instead, i.e., allowing the transition-transversion bias to vary along the sequence and relying on the F3x4 codon substitution model, had a considerable effect on the denominator of dS, but did not change any qualitative difference (i.e., the ordering of the comparisons). In humans, we used the low coverage pilot data from the 1000 Genomes YRI (30); however, instead considering data generated by the Seattle SNP project (http://pga.gs.washington.edu/) using Sanger sequencing yielded highly similar results.

The 95% confidence intervals for dS were calculated as follows: we estimated the mean dS in 585 bps for the A alleles using *codeml* with the species tree specified, then divided it by the length of the tree to obtain the expected number of mutations per bp per My. We did the same for the B alleles, and took the mean of the estimates from the two trees, $m$, which was 0.0013/My/bp in Hominoids and 0.0025/My/bp in Old World Monkeys. We then assumed that synonymous mutations are neutral, so that the number of synonymous mutations in a lineage is Poisson distributed with mean $\lambda$, where $\lambda=mT^*L$, $T^*$ is the divergence time between lineages from different species and $L$ is the number of synonymous sites (estimated by PAML). $T^*$ values were taken from previously published estimates, notably from ref. (29): 20 My for human-gibbon, 18 My for colobus-macaque, 17 My for human-orangutan, 12 My for vervet-macaque, 8 My for baboon-macaque. For African apes, a synonymous allele (G at nucleotide 813) is fixed in human and gorilla but absent from chimpanzee, suggesting that gorilla is closer to human than chimpanzee/bonobo in this region, i.e., indicating a case of incomplete lineage sorting (50). If so, then lineages from humans, chimpanzees and gorillas coalesced before the split of the three species. In order to account for this possibility, we used 8 My as the divergence time between human, chimpanzee and gorilla lineages. This is conservative for our purposes, as a more recent divergence would imply a lower 95% CI, and an even greater signal of trans-species polymorphism.

*Test of convergent evolution using the internal branches in the Hominoid tree.* Using the 585 bps sequence for which we have data for all species, we inferred where synonymous substitutions occurred along the tree of Hominoids by parsimony, verifying that PAML output yielded similar ancestral sequences (using the maximum likelihood method based on an *a priori* species tree, with gorilla closest to human). We then calculated the rate of synonymous substitutions on extant lineages that are monomorphic (chimpanzee, gorilla and siamang) or polymorphic (human and gibbon). For this purpose, we ignored the bonobo lineage, since the addition of the bonobo lineage (which split from chimpanzee < 1 Mya 56) does not provide much time for neutral mutations to arise and fix, so is relatively uninformative. In addition, the orangutan lineage was excluded from this analysis because it appears to have experienced a recent turnover.

We assumed that B arose ~8 Mya (the minimum age indicated by Fig 2b) and tested a null model in which there was only one *ABO* class during the first 12 My of Hominoid evolution. Specifically, we compared rates of substitutions in exon 7 on internal branches (1 substitution in 24 My) to that seen in monomorphic lineages (6 substitutions in 24 My). Assuming that the number of synonymous fixations is Poisson distributed, this yields *p*-value = 0.01. In contrast, when we asked if the internal branches differ from extant branches that are polymorphic (2 mutations in 16 My), we could not reject the null model: *p*-value = 0.18. Comparing the two tests, the posterior odds for a model of trans-species polymorphism rather than convergent evolution are 10:1. Considering instead that this region is not a case of incomplete lineage sorting among African apes (i.e., that the site at nucleotide 813 experienced a recurrent mutation in human and gorilla) does not change the conclusions (in this case, the posterior odds are 34:1). A similar test was not performed in Old World Monkeys because we do not have sequence data for monomorphic species from which to estimate the rate of synonymous divergence.

We note that this test assumes (conservatively for our purposes) that lineages polymorphic at present were always polymorphic and lineages monomorphic at present lost the polymorphism immediately after speciation. Beyond requiring that the balanced polymorphism be stably maintained in polymorphic lineages, it does not make assumptions about the strength or mechanism of selection.

## Acknowledgements

We thank P. Andolfatto, G. Coop, R. Hudson, G. Perry, J. Pritchard and M. Stephens for helpful discussions, and G. Coop and J. Pritchard for comments on an earlier version of the manuscript. This study used biological materials obtained from the Southwest National Primate Research Center, which is supported by NIH-NCRR grant P51 RR013986. The complete source of all primate materials is described in SOM. This work was supported by a Rosalind Franklin Award and R01 GM72861 to MP. CO was partially funded by the grant R01 HD21244. EET was supported by the grant K12 HL090003. GS was funded by a Flegg fellowship and Israel Science Foundation grant (no. 1492/10). M.P. is a Howard Hughes Early Career Scientist.

## Author Contributions

EET, KG, JL, SWM and CO initiated the study, which was extended by LS, EET, GS, MP and CO. KG, JM, SWM and SR provided samples. EET, JL and LS collected the sequence data. LS, EET, TF, PA, AV, GS and MP analyzed the data. LS, GS, CO and MP wrote the paper.

## Author Information

All sequences produced for this study have been deposited in Genbank (JQ857042-JQ857076). The authors declare no competing financial interests.

**Figure Legends**

Figure 1: The phylogenetic distribution of ABO phenotypes and genotypes

Shown is a phylogenetic tree of primate species, with a summary of phenotypic/genotypic information given in the first column, and the genetic basis for the A versus B phenotype provided in the second column (in capital letters are the functionally important codons at positions 266 and 268). See Supplementary Table 1 for the source of information about phenotypes/genotypes. Only species with available divergence times are represented here (34 out of 40). The phylogenetic tree is drawn to scale, with divergence times (on the x axis) in Millions of years taken from ref. (29). The abbreviation OWM refers to Old World Monkeys and NWM to New World Monkeys. Under a model of convergent evolution, these data suggest that A is the ancestral allele, and a turnover (e.g., a neutral substitution) occurred on the branch leading to Old World Monkeys. If instead B were ancestral, all Old World Monkeys would have had to serendipitously converge from ATG to TTG to encode a Leucine, while all New World Monkeys and Hominoids converged to the CTG codon.

Figure 2: Expectations under the two possible evolutionary hypotheses

Schematic of expectations for (a) the trans-species polymorphism and (b) the convergent evolution hypotheses. We illustrate the latter case assuming that A is ancestral. The population tree is outlined in gray, and the A and B lineages sampled from three species are shown in blue and red, respectively. Under the trans-species polymorphism hypothesis, the divergence between A and B is deeper than the species split and species share a short ancestral segment with shared polymorphisms; neither of these patterns is expected under a model of convergent evolution.

(c) Plot of the expected length of the segment in which a signal of a trans-species polymorphism should be detectable. Specifically, we present the expected (two-sided) segment contiguous to the selected site in which the divergence between A and B lineages from different species should exceed the divergence between A (or B) lineages from different species (see Methods). We considered a recombination rate of 1 cM/Mb, as the recombination rate in this exon is estimated to be ~1 cM/Mb in humans (57) and higher in chimpanzees (58), and varied the generation time between 10 and 20. Higher recombination rates lead to shorter segment lengths (see Supplementary Figure 1).

Figure 3: Tree of *ABO* exon 7 alleles in (a) Hominoids and (b) Old World Monkeys.
In (a), the tree is based on 300 bps; in (b), the tree is based on a smaller window of 200 bps, because the shorter generation time in Old World Monkeys should lead to a smaller segment with a signal of a trans-species polymorphism (2) (see Figure 2c and Methods). The tree is centered on the two functional sites. We excluded rare recombinant haplotypes between functional classes (see Methods). In red is the median clade credibility (based on three runs that yielded identical consensus trees), i.e., the proportion of trees sampled from the posterior distribution that had this clade (51).

Figure 4: Evidence for a trans-species polymorphism in Old World Monkeys and Hominoids. Shown in (a) for Old World Monkeys and (b) for Hominoids are the synonymous pairwise differences (dS) among *ABO* haplotypic classes in 201 bps (i.e. 67 codons) sliding windows, as well as the shared SNPs between the species compared. For details about how dS and the 95% confidence interval were estimated, see Methods. Genome-wide mean synonymous diversity estimates within species were taken from ref. (46). When multiple species were available per genera, we chose one representative with the largest sample size, namely *Macaca mulatta*, *Colobus Angolensis* and *Hylobates Lar*. Only the most informative comparisons are presented here, with the rest shown in Supplementary Figures 2 and 3 (which also includes similar figures using a larger sliding window choice of 300 bps). Because of recombination, the segment carrying the footprint of a trans-species polymorphism will not necessarily be contiguous and hence, even if there were no stochasticity in the mutation process, diversity levels may be jagged and may only be unusually deep in (at most) small windows (2).
In panel (c) is a depiction of the synonymous substitutions inferred to have occurred in the Hominoid phylogeny (represented as ticks, in pink for monomorphic lineages with only one allelic class, in purple for polymorphic lineages with multiple allelic classes). Numbers along the lineages represent Millions of years. The A and B lineages are shown in blue and red, respectively. Orangutan is not shown because it appears that a recent turnover occurred in this species, so the lineage is not informative for our test; similarly, the branch from the common ancestor with chimpanzee to bonobo is too short to be informative (see Methods). Parsimony was used to assign synonymous changes to Hominoid lineages. Gorilla is shown closest to human because a substitution (at a non-CpG site) is inferred to have occurred in the ancestor of humans and gorillas but is not found in chimpanzees, suggesting incomplete lineage sorting in this region of the genome (50) (Supplementary Table 2); treating it instead as a multiple hit in humans and gorillas only strengthens our conclusions.

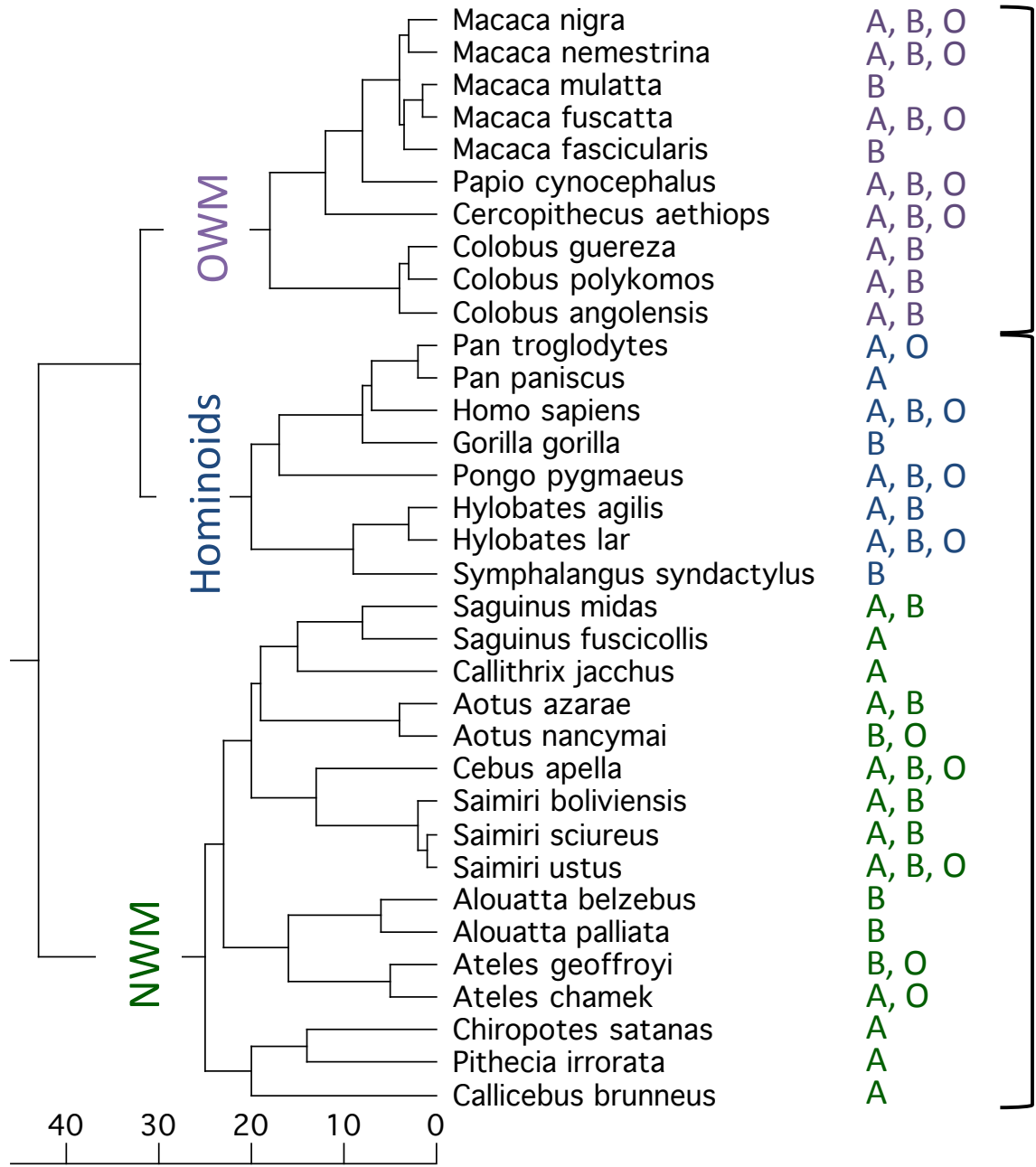

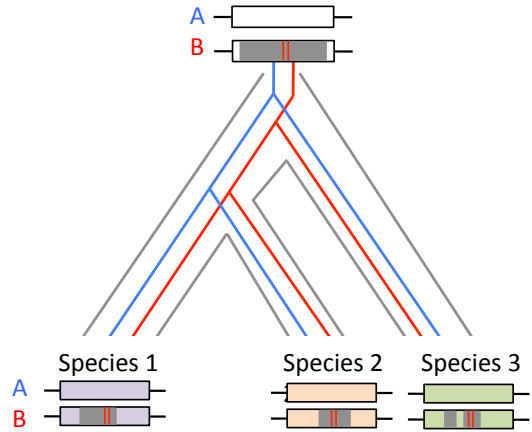
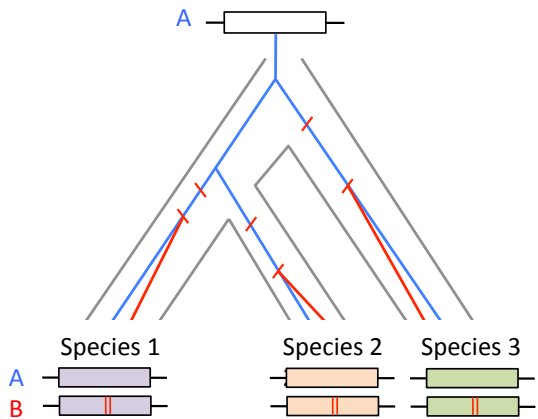
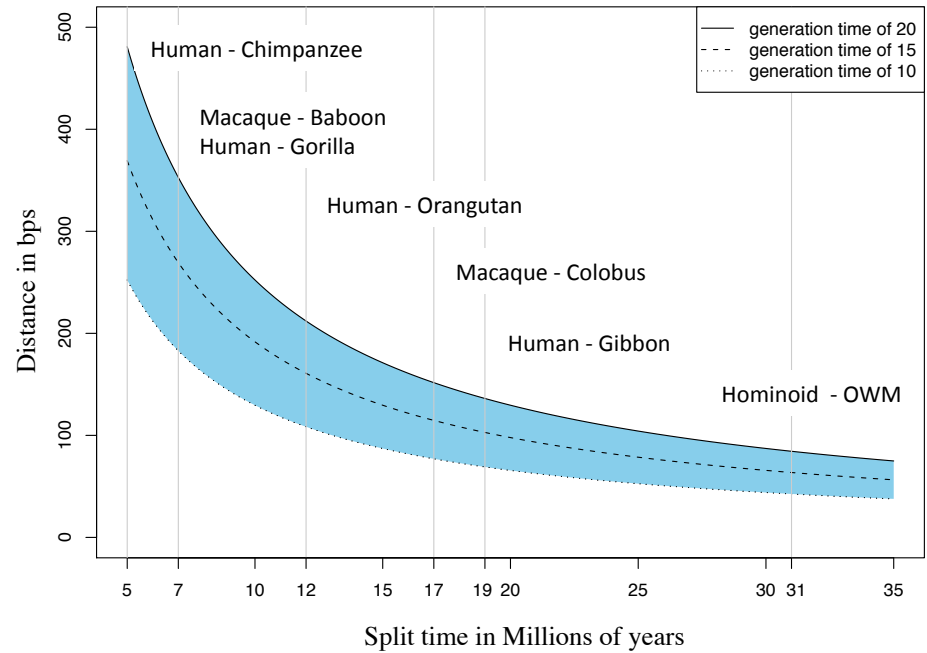

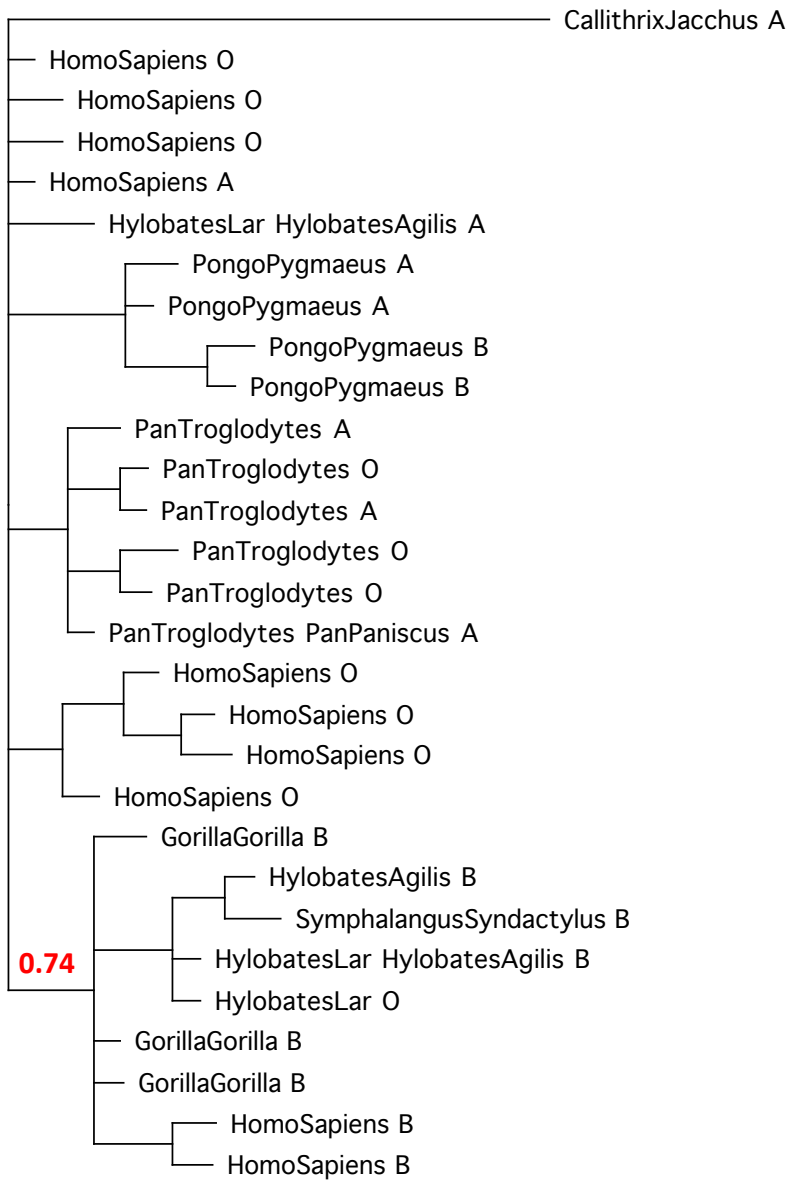
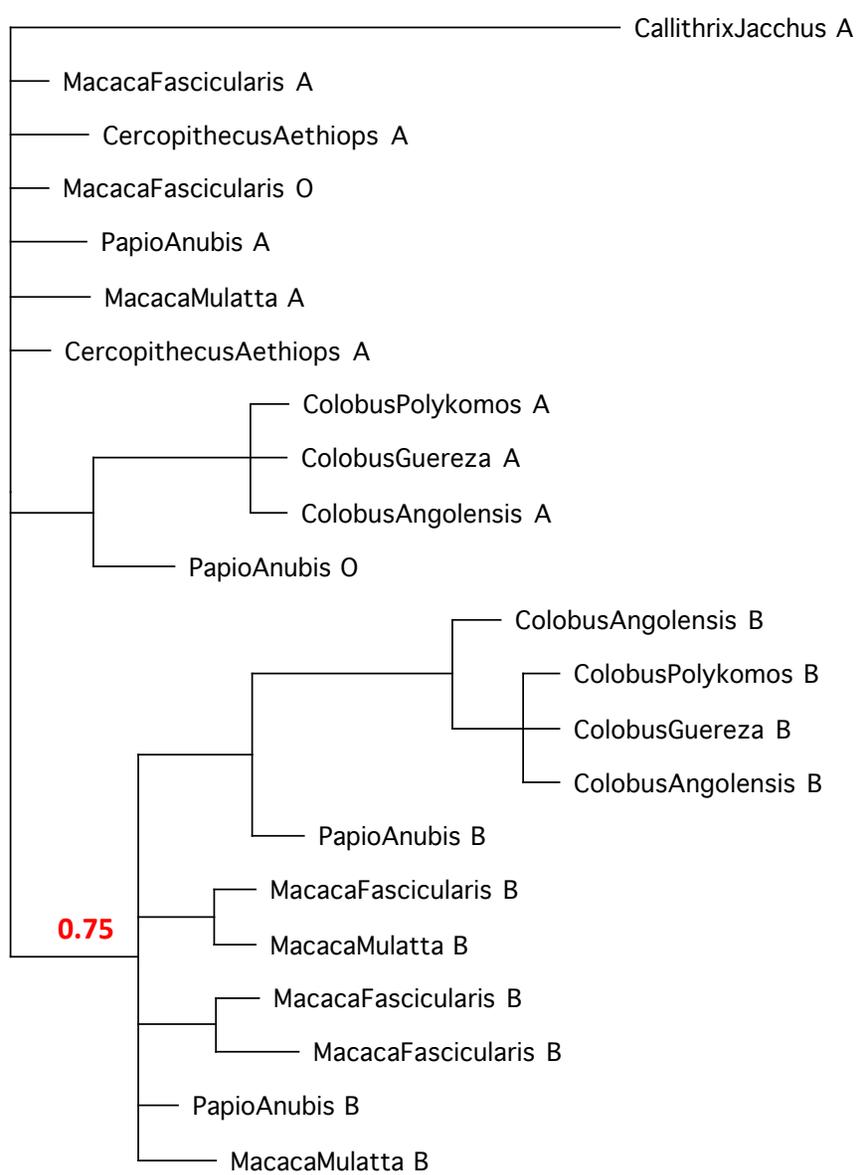

a. Hominoids

b. Old World Monkeys

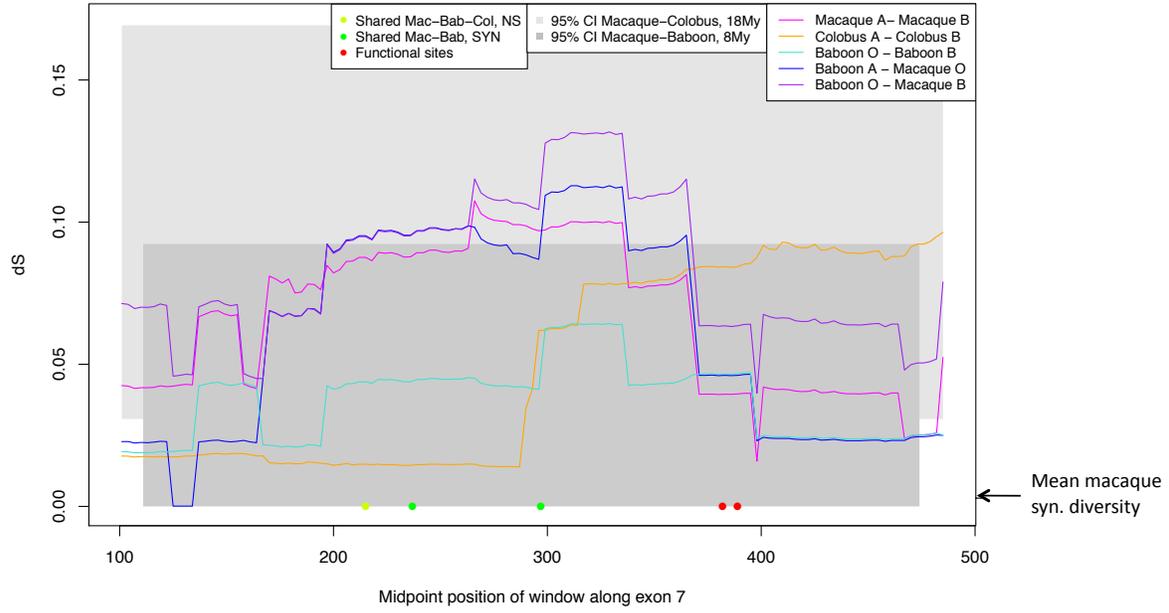

a.

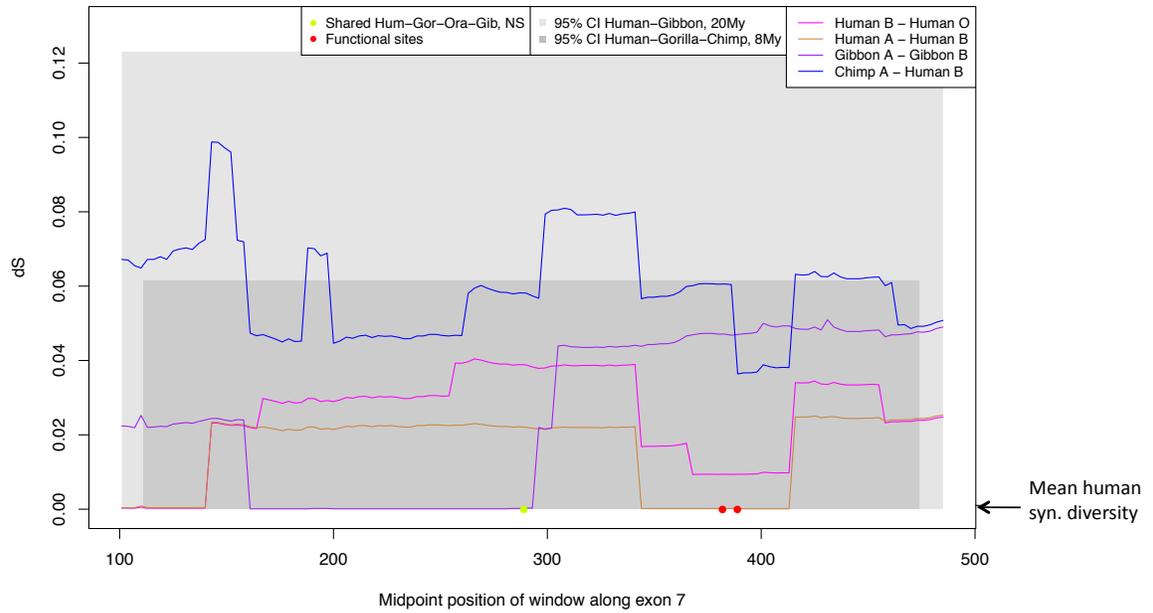

b.

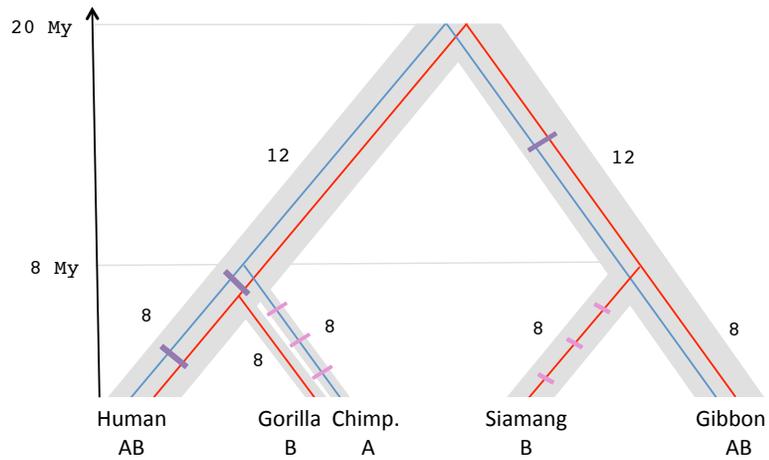

c.

**Supplementary Material**

**Supplementary Legends**

**Supplementary Figure 1**: Sensitivity of the expected segment length to the recombination rate and effective population size. The derivation is described in the Methods; r is the recombination rate per bp per generation and N the effective population size. For the sensitivity to the generation time, see Figure 2c. For other details, see legend of Figure 2c.

**Supplementary Figure 2**: Synonymous pairwise differences (dS) among ABO alleles in Old World Monkeys: (a) in macaques, (b) in colobus monkeys and (c) in baboons, for 201 bps (left panel) and 300 bps (right panel) sliding windows.
For further details, see the legend of Figure 4a-b and Methods. We note that erosion of the ancestral segment by recombination can lead to divergence times that are much more recent than the origin of the ABO polymorphism (see main text and ref. (1)). Thus, deeper divergence levels than expected are evidence for a trans-species polymorphism, but shallow divergence levels are not evidence against it.

**Supplementary Figure 3**: Synonymous pairwise differences (dS) among ABO alleles in Hominoids: (a) in humans, (b) among human, chimpanzee and gorilla, (c) in gibbons and (d) in orangutans, for 201 bps (left panel) and 300 bps (right panel) sliding windows.
Other details are as in Supplementary Figure 2.

**Supplementary Figure 4**: Plot of the expected length of the segment in which a signal of a trans-species polymorphism should be detectable. Shown are the relevant scale for inter-species comparisons (Figure 4a-b) and the one for the tree (Figure 3) (see Methods). For other details, see legend of Figure 2c.

**Supplementary Table 1**: Phenotype and genotype studies of ABO in primates.
Included are all 40 non-human species with phenotype or genotype information for ABO in the literature. We did not consider phenotypic studies with fewer than five individuals (except for siamang, to bolster the evidence that they are monomorphic for B). When available, the observed phenotypes are given; the exception is when only the allele frequencies were reported, in which case the alleles are highlighted in blue. Lines in grey represent species without information about divergence times in ref. (2) (therefore not shown on the tree in Figure 1) or cases in which ABO phenotypes are only given for a combination of species.

*In gorillas, phenotypic tests using red blood cells are unreliable, as gorillas have few (if any) antigens at the surface of these cells. Gorillas were therefore classified based on their saliva or serum phenotype (see 3, 4) as well as based on their genotype, both of which consistently categorize them as monomorphic for the B allele.

**Supplementary Table 2**: Table of substitutions and polymorphisms that require more than one mutation given the species tree: ((((((Human, Gorilla), Chimpanzee), Orangutan), (Gibbon, Siamang)), (((Baboon, Macaque), Vervet), Colobus)), (Marmoset, Howler)).
Parsimony was used to assign substitutions to lineages. Each row represents a unique haplotype; n denotes the number of chromosomes surveyed in the subspecies/species (left column). The two functional sites are shown in red. Identical polymorphisms shared between two or more species are shown in green; these could arise from mutations on an ancestral segment shared by descent or from recurrent mutations. For comparison, non-identical shared polymorphisms, which reflect recurrent mutations, are shown in blue. Nucleotide positions start at the first ATG codon in the mRNA sequence (Genbank NM_020469.2). We did not include SNPs shared among species of macaques, species of colobus monkeys or species of gibbons, because they are more recently diverged and therefore more likely to share SNPs because of neutral incomplete lineage sorting (5). The sequence data for yellow baboon are from ref. (6); they are not analyzed elsewhere because they are not a population sample.
*cis-AB allele (having both A and B activity).
We note that, consistent with old tMRCAs between *ABO* alleles, a SNP is shared between humans, gorillas, orangutans and gibbons on the B background. This non-synonymous SNP (amino acid 235) is known to affect ABO activity in humans (7), so may denote a subclass of B alleles.

**Supplementary Table 3**: Number of polymorphic sites found per species, both for synonymous and non-synonymous sites, along with the total number of polymorphic sites across the dataset (i.e., polymorphic in at least one species).
n: Number of individuals; S: number of polymorphic sites; Syn: number of synonymous polymorphic sites; NSyn: number of non-synonymous polymorphic sites.

**Supplementary Table 4**: Primer sets used for amplification and sequencing of *ABO* exon 7. Degenerate positions are given according to the IUPAC code (S=G/C; R=A/G; K=G/T).

**Supplementary Table 5**: Polymorphic sites in *ABO* exon 7 found in chimpanzee, gorilla and orangutan.

Data from previous studies and this one are summarized. Grey boxes represent polymorphic sites, and orange boxes highlight the inconsistencies between ref. (8) and other studies. The nomenclature of each allele is as it was in the original study. Nucleotide positions were taken from the mRNA sequence, starting at the first ATG (nucleotide sequence NM_020469.2 in Genbank). As can be seen, the data from ref. (8) appear to be unreliable.

**Supplementary Notes**

**1- Summary of previous studies about ABO evolution**

Phylogenetic approaches have been used to show that, in species of gibbons and siamangs (9) as well as in species of macaques (10-12), *ABO* haplotypes cluster by allelic type rather than by species. Additionally, in gibbons/siamangs, coalescence times between the A and B alleles were estimated to be older than split time between the Hylobates and Symphalangus genera. The case of baboon has led to more controversy, with ref. (10) finding that macaques and baboons shared the same trans-species (TS) polymorphism, whereas refs. (11) and (12) reported that olive and yellow baboons form a monophyletic cluster as compared to macaques. Thus, among recently diverged subspecies or species (i.e., among gibbons/siamangs and among macaques), there is agreement that the data support a trans-species polymorphism, but not for more distant species.

Ref. (8) is the only study to have favored the TS polymorphism hypothesis among other Hominoids. Their conclusion was based on the phylogenetic tree of two A chimpanzees, five B gorillas and one A orangutan, which used 405 bp sequences of exon 7. When we compared the sequences for chimpanzee, gorilla and orangutan obtained by different authors (6, 8, 11, 13), as well as by this study, however, we found that the dataset from ref. (8) appears to contain sequencing errors (see Supplementary Table 5).

All other studies concluded that ABO sharing results from convergent evolution. For example, ref. (14) obtained a phylogenetic tree matching the species tree for human, chimpanzee, gorilla, orangutan, baboon and macaque. In order to obtain this tree, however, they assumed that all nucleotides shared between species were due to parallel substitutions. Using a neighbor-joining method, ref. (11) also obtained a tree supporting the independent evolution of the B allele in humans, gorilla and orangutan, but with insignificant bootstrap support.

In addition, refs. (14) and (15) estimated divergence times among human ABO alleles to be between 2.7-4.7 Millions of years (Myr) (based on 405 bp in exon 7) and 4.8 Myr (based on ~2.5kb covering exon 6 and 7), respectively, and interpreted these numbers as evidence that the B allele originated more recently than the human-chimpanzee split. Ref. (16), using data from intron 5 and intron 6 in three humans, two chimpanzees and two gorillas, did not find any shared polymorphism between species and, on that basis, argued against a TS polymorphism. Similarly, refs. (6) and (10) considered the existence of ten substitutions differentiating OWM from Hominoids as evidence that the A and B alleles of OWM appeared independently from those of Hominoids. However, these arguments implicitly ignore the effect of recombination, which restricts the presence of shared SNPs to only a small region around the functional sites, allowing for the accumulation of substitutions between OWM and Hominoids outside of this region (as pointed out by refs. 1, 17) and leading to a decreased estimate of the divergence time of ABO alleles if too large a window is considered (18).

**2- Additional sample information**

The authors appreciate general support from Species Survival Plan® coordinators and veterinary advisors (Dr. Gay Reinartz and Dr. Victoria Clyde, Bonobo SSP; Dr. Kay Backues, Common Chimpanzee SSP; Dr. Dan Wharton and Dr. Tom Meehan, Gorilla SSP; Lori Perkins and Dr. Rita McManamon and Dr. Chris Bonar, Orangutan SSP), and Dr. Kristin Lukas, Dr. Elizabeth Lonsdorf and Maureen Leahy. Contributing institutions are acknowledged for their provision of samples and animal identification, as described in ref. (19): *Species Survival Plan*® Busch Gardens Tampa Bay, Tampa, FL; Columbus Zoo and Aquarium, Powell, OH; Denver Zoological Gardens, Denver, CO; Fort Wayne Children's Zoo, Fort Wayne, IN; Gladys Porter Zoo, Brownsville, TX; Honolulu Zoo, Honolulu, HI; Houston Zoo, Inc., Houston, TX; Jacksonville Zoo and Gardens, Jacksonville, FL; Lincoln Park Zoo, Chicago, IL; The Maryland Zoo in Baltimore, Baltimore, MD; Memphis Zoo, Memphis, TN; Miami Metrozoo, Miami, FL; Riverbanks Zoo & Garden, Columbia, SC; San Diego Zoo's Wild Animal Park, Escondido, CA; Sedgwick County Zoo, Wichita, KS; Smithsonian National Zoological Park, Washington, DC; Utah's Hogle Zoo, Salt Lake City, UT; Zoo New England, Stoneham, MA.

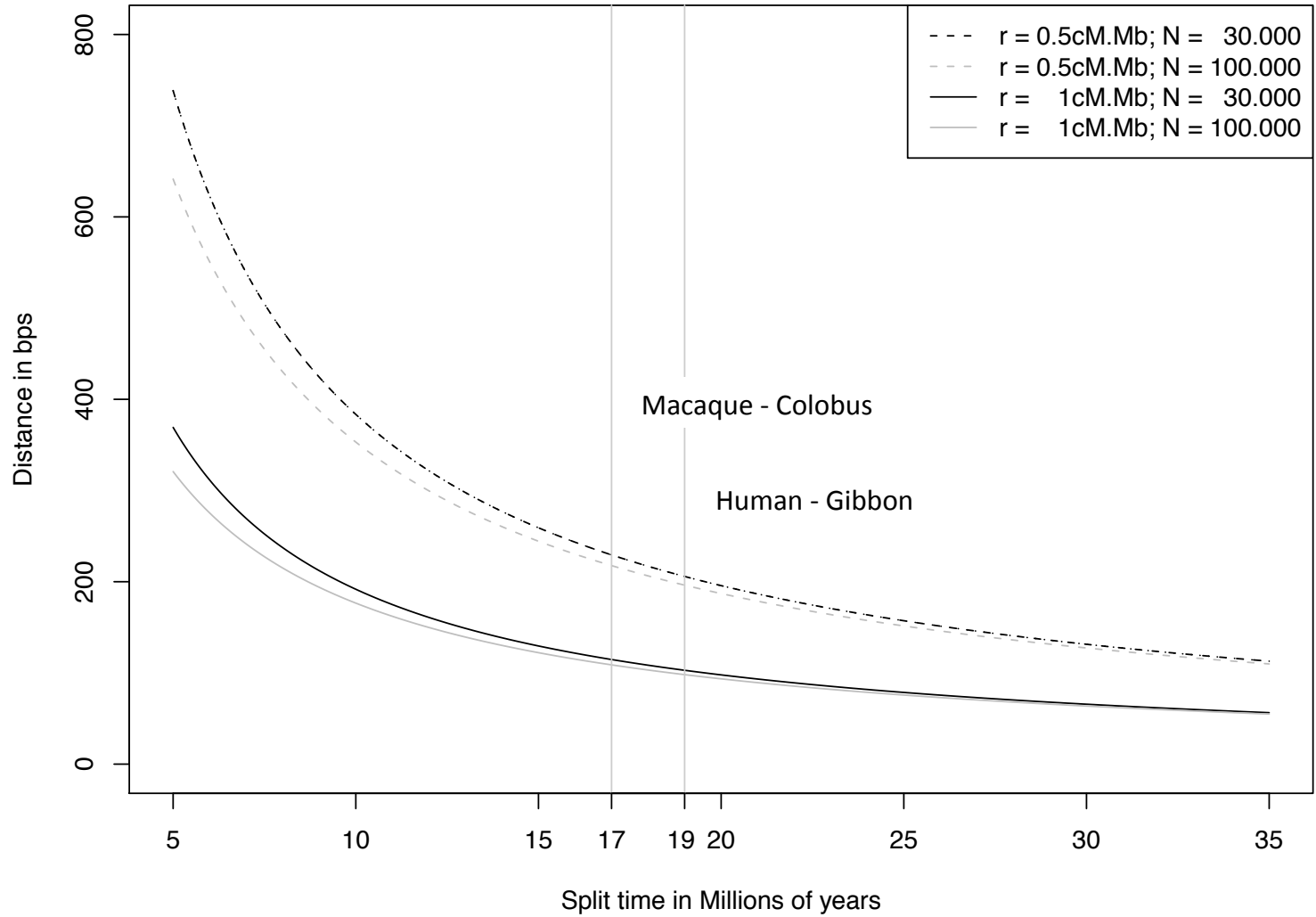

# 201bp                                                                    300bp

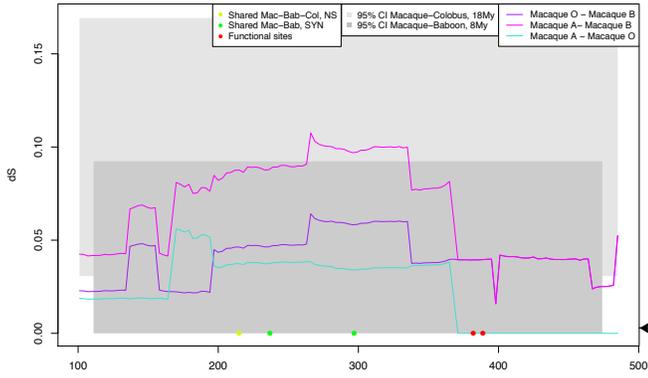
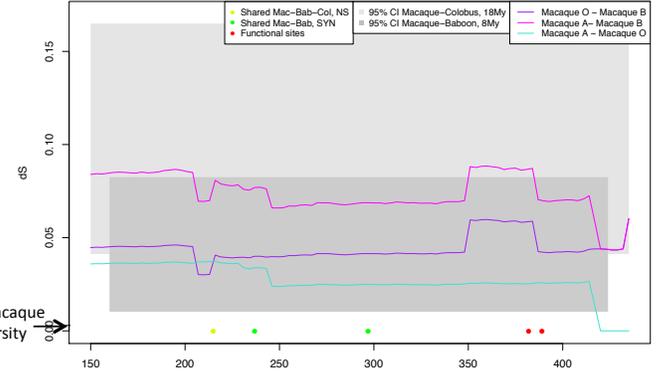
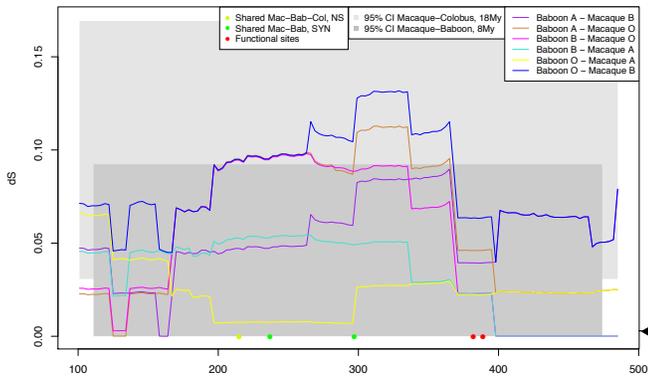
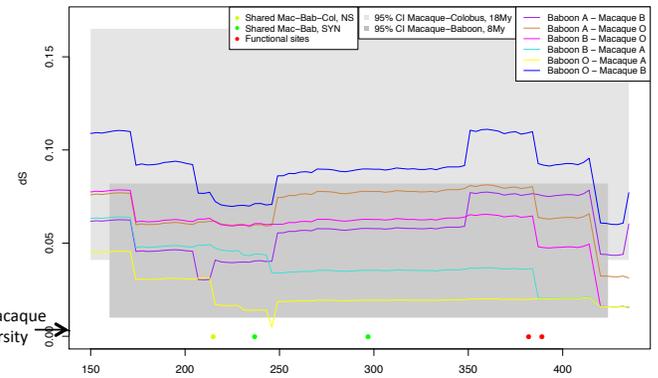
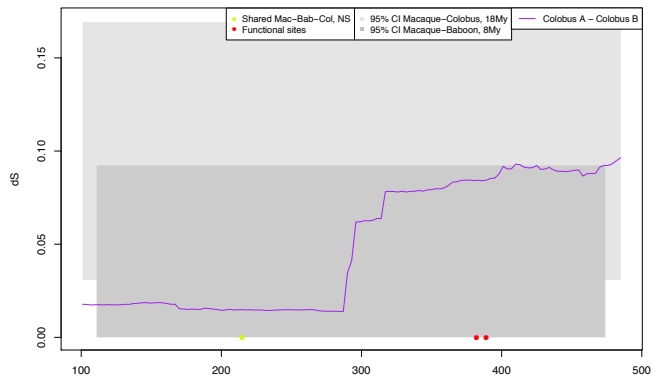
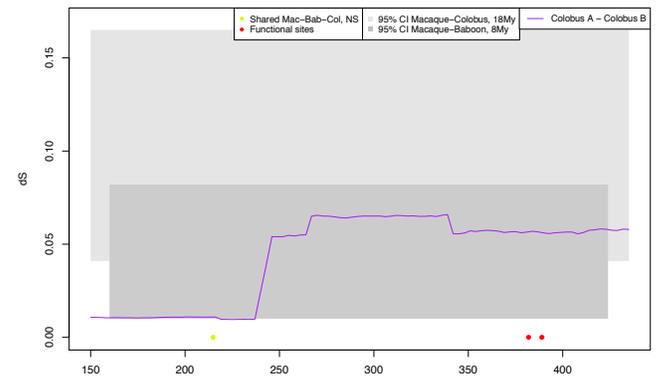
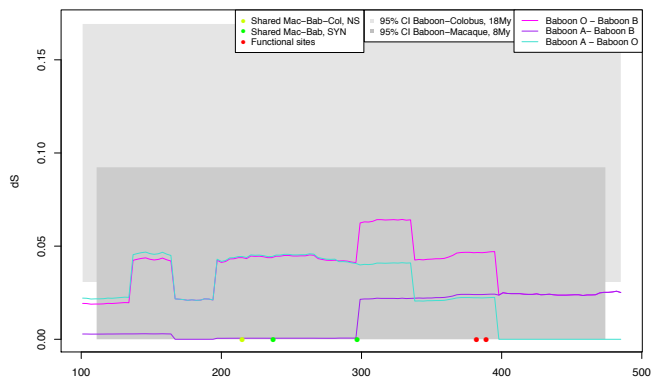
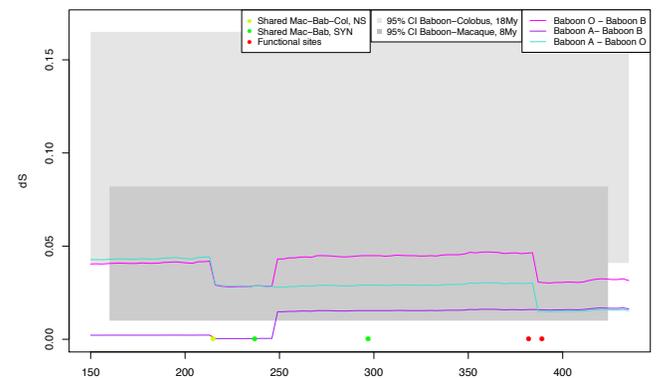

## 201bp

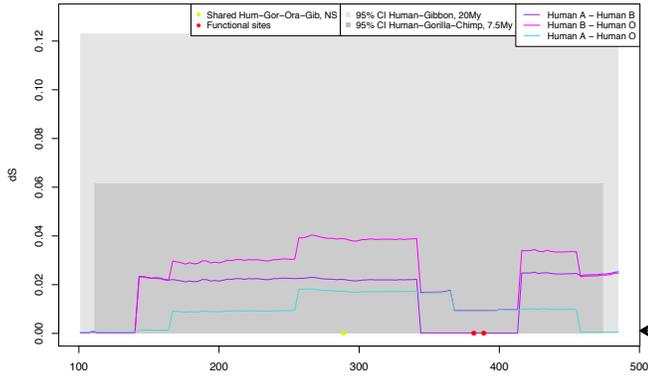

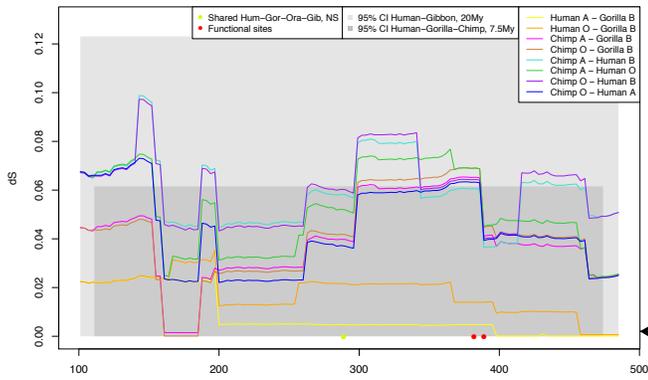

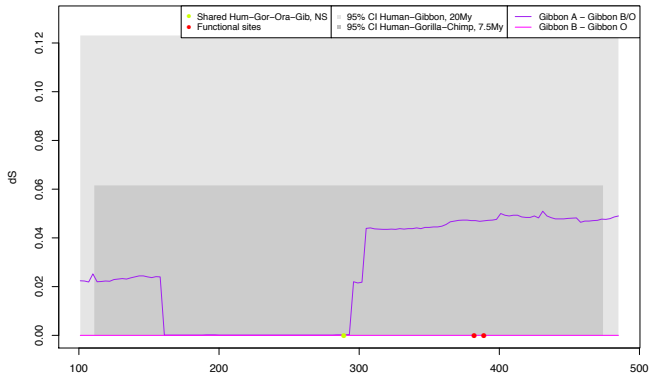

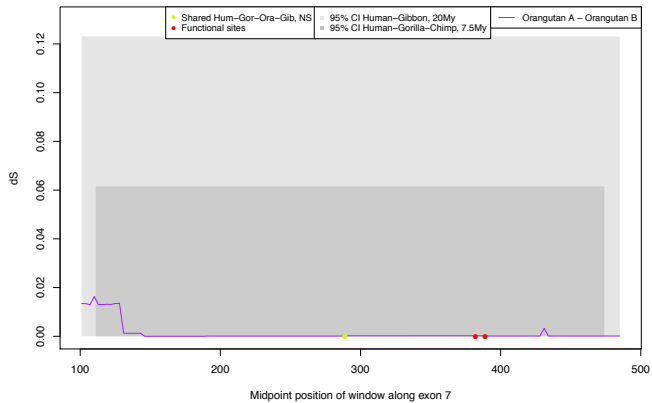

## 300bp

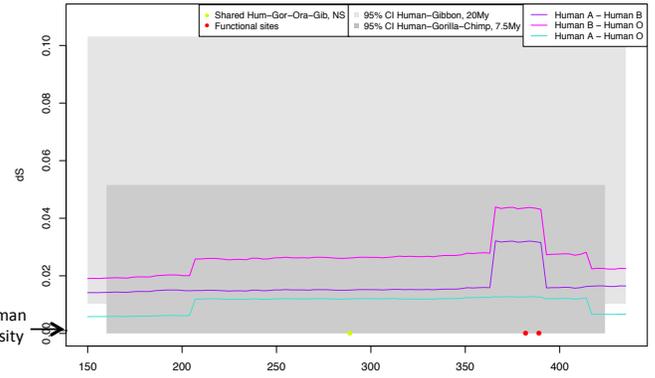

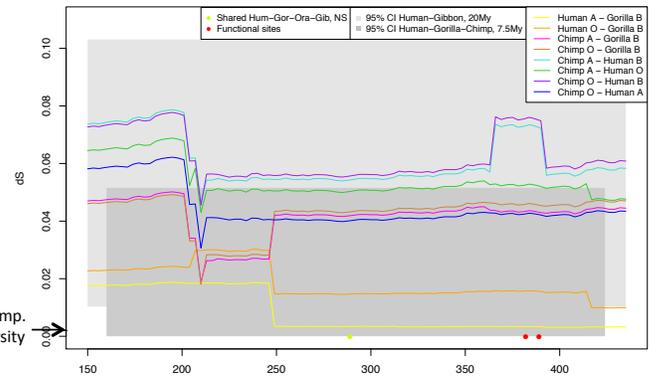

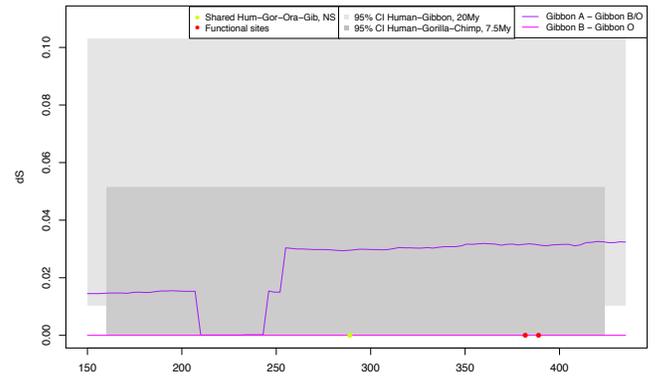

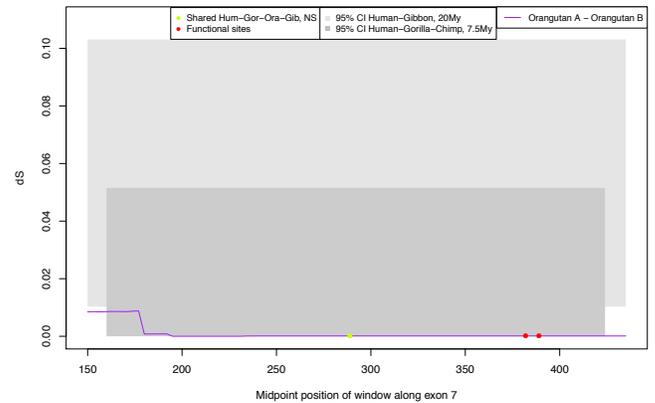

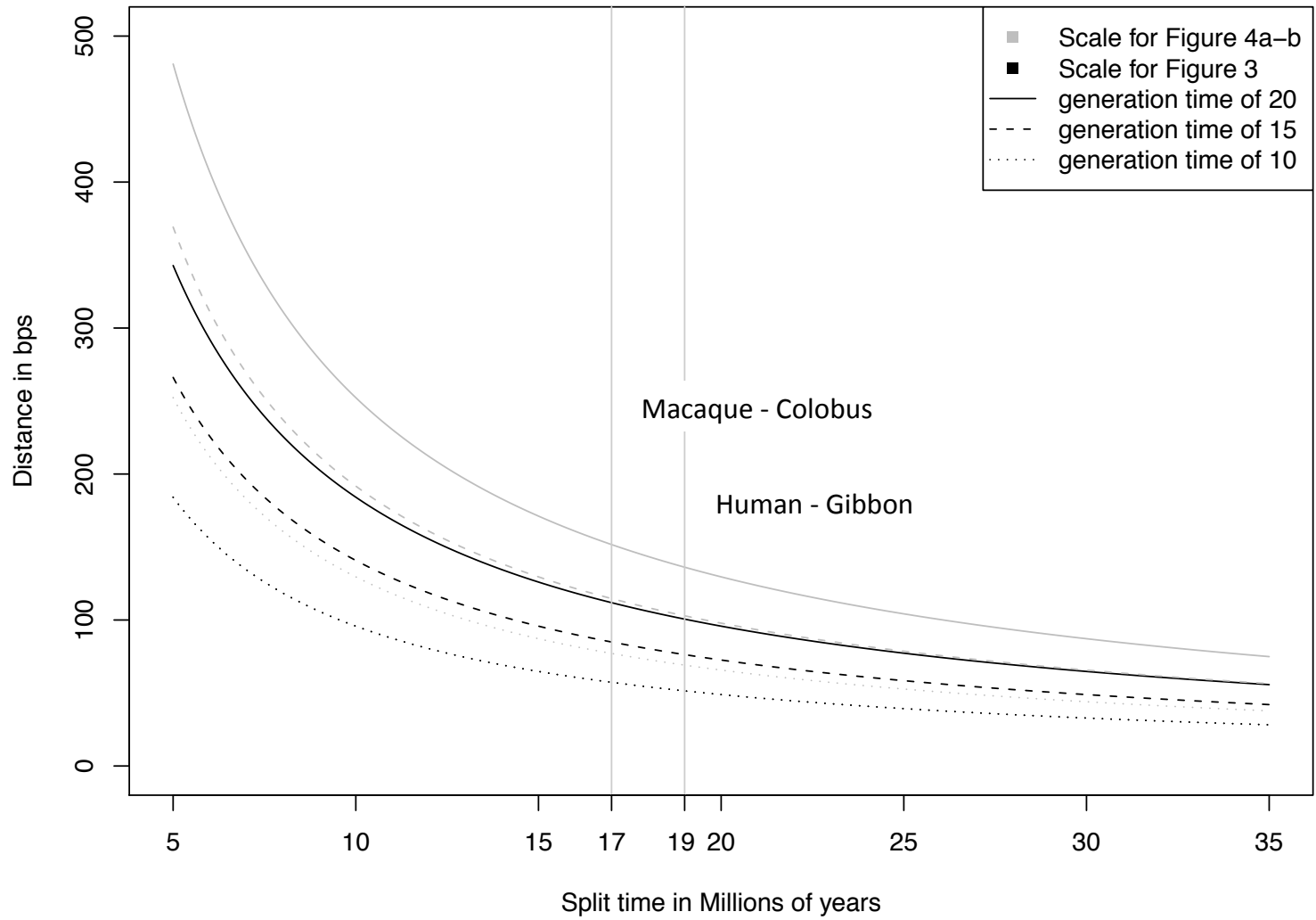

| | Common name | Species name | Phenotype | N ind | Reference | Genotype | N ind | Reference |
|---|---|---|---|---|---|---|---|---|
| **New World Monkeys** | Marmoset | Callithrix jacchus | A | 45 | Socha & Ruffie, 1983 | A | 4 | This study |
| | | Callithrix jacchus jacchus | A | 35 | daRocha et al, 1990 | | | |
| | | Callithrix emiliae | A | 47 | daRocha et al, 1990 | | | |
| | Tamarins | Saguinus midas niger | A, B, AB | 81 | Schneider et al, 1987 | | | |
| | | Saguinus fuscicollis | A | 18 | Wiener et al, 1967 | | | |
| | | Saguinus fuscicollis weddeli | A | 88 | daRocha et al, 1990 | | | |
| | | Saguinus nigricollis | A | 8 | Gengozian, 1964 | | | |
| | | Saguinus nigricollis | A | 6 | Wiener et al, 1967 | | | |
| | Capuchins | Cebus apella | A, B, O | 95 | Corvelo et al, 1987 | | | |
| | | Cebus apella | A | 5 | Socha & Ruffie, 1983 | | | |
| | | Cebus apella | A, B, AB, O | 96 | Corvelo et al, 2002 | | | |
| | | Cebus apella | A, O | 130 | daRocha et al, 1992 | | | |
| | | Cebus apella apella | A, B, AB, O | 19 | Schneider et al, 1985 | | | |
| | | Cebus apella paraguyanus | A | 55 | HaradaHamel et al, 1988 | | | |
| | Squirrel Monkey | Saimiri (multiple) | A, AB, O | 11 | Socha & Ruffie, 1983 | | | |
| | | Saimiri ustus | A, B, AB, O | 117 | daRocha et al, 1992 | | | |
| | | Saimiri boliviensis boliviensis | A, AB | 76 | Schneider et al, 1991 | | | |
| | | Saimiri boliviensis peruviensis | A, AB, B | 44 | Schneider et al, 1993 | | | |
| | | Saimiri boliviensis peruviensis | A, B | 22 | Terao et al, 1988 | | | |
| | | Saimiri boliviensis peruviensis | A, B, O | 17 | Terao et al, 1988 | | | |
| | | Saimiri boliviensis peruviensis | A, B, O | 18 | Terao et al, 1988 | | | |
| | | Saimiri sciureus | A, AB | 74 | Corvelo et al, 2002 | | | |
| | | Saimiri sciureus macrodon | A, AB | 8 | Schneider et al, 1993 | | | |
| | Night monkey | Aotus azarae boliviensis | B, AB | 84 | daRocha et al, 1992 | | | |
| | | Aotus nancymai | B, O | 93 | HaradaHamel et al, 1990a | | | |
| | | Aotus vociferans | B, O | 20 | HaradaHamel et al, 1990a | | | |
| | | Aotus infulatus | B | 40 | Corvelo et al, 2002 | | | |
| | Saki Monkey | Chiropotes satanas utahicki | A | 72 | HaradaHamel et al, 1990b | | | |
| | | Pithecia irrorata irrorata | A | 117 | daRocha et al, 1992 | | | |
| | Titi Monkey | Callicebus brunneus | A | 130 | daRocha et al, 1992 | | | |
| | Howler Monkey | Alouatta belzebus | B | 50 | Corvelo et al, 1985 | | | |
| | | Alouatta palliata | B | 52 | Froehlich et al, 1977 | | | |
| | | Alouatta caraya | | | | B | 3 | This study |
| | Spider Monkey | Ateles (paniscus) chamek | A, O | 31 | daRocha et al, 1992 | | | |
| | | Ateles (paniscus) chamek | A | 6 | Wiener et al, 1942 | | | |
| | | Ateles geoffroyi grisescens | B, O | 5 | Wiener et al, 1942 | | | |
| **Old World Monkeys** | Vervet | Cercopithecus aethiops | A, B, AB, O | 330 | Dracapoli & Jolly, 1983 | A | 4 | This study |
| | Macaque | Macaca fascicularis | A, B, AB, O | 729 | Malaivijitnond et al, 2007 | A, B, O | 7 | Doxiadis et al, 1998 |
| | | Macaca mulatta | A, B, AB, O | 160 | Malaivijitnond et al, 2007 | A, B | 6 | Doxiadis et al, 1998 |
| | | Macaca mulatta | B | 6 | Moor-Jankowski et al, 1964 | | | |
| | | Macaca nemestrina | B | 5 | Moor-Jankowski et al, 1964 | | | |
| | | Macaca nigra | A, B, O | 10 | Moor-Jankowski et al, 1964 | | | |
| | | Macaca irus | A, B, AB, O | 12 | Wiener & Moor-Jankowski, 1963 | | | |
| | | Macaca fuscatta | | | | B | 5 | Noda et al, 2000 |
| | Baboon | Papio anubis | A, B, AB, O | 31 | Diamond et al, 1997 | A, B, O | 31 | Diamond et al, 1997 |
| | | Papio anubis + cynocephalus | A, B, AB | 76 | Wiener & Moor-Jankowski, 1963 | | | |
| | | Papio anubis + cynocephalus | A, B, AB | 11 | Moor-Jankowski et al, 1964 | | | |
| | | Papio cynocephalus | | | | A, B | 2 alleles | Kominato et al, 1992 |
| | Colobus | Colobus angolensis | | | | A, B | 3 | This study |
| | | Colobus polykomos | | | | A, B | 1 | This study |
| | | Colobus guereza | | | | A, B | 1 | This study |
| **Hominoids** | Chimpanzee | Pan troglodytes | A, O | 91 | Wiener & Moor-Jankowski, 1963 | | | |
| | | Pan troglodytes | A, O | 39 | Moor-Jankowski et al, 1964 | A | 35 | This study |
| | | Pan troglodytes | A, O | 234 | Kermarrec et al, 1999 | | | |
| | | Pan troglodytes | A, O | 233 | Gamble et al, 2010 | | | |
| | Bonobo | Pan paniscus | A | 71 | Gamble et al, 2010 | A | 10 | This study |
| | | Pan paniscus | A | 5 | Moor-Jankowski et al, 1975 | | | |
| | Gorilla (lowland) | Gorilla gorilla gorilla | B, O, A* | 152 | Gamble et al, 2010 | B | 31 | This study |
| | | Gorilla gorilla gorilla | B | 27 | Socha & Moor-Jankowski, 1979 | | | |
| | Gorilla (mountain) | Gorilla gorilla beringei | B | 6 | Socha & Moor-Jankowski, 1979 | | | |
| | Orangutan (Borneo) | Pongo pygmaeus pygmaeus | A, B, AB, O | 169 | Gamble et al, 2010 | A, B | 12 | This study |
| | | Pongo pygmaeus pygmaeus | A, B, AB | 16 | Moor-Jankowski et al, 1964 | | | |
| | | Pongo pygmaeus pygmaeus | A, AB | 10 | Wiener & Moor-Jankowski, 1963 | | | |
| | Orangutan (Sumatra) | Pongo pygmaeus abelii | A, B, AB | 33 | Gamble et al, 2010 | | | |
| | Gibbon | Hylobates agilis | A, B, AB | 140 | Socha & Moor-Jankowski, 1979 | A, B | 5 | Kitano et al, 2009 |
| | | Hylobates lar | | | | A, B, O | 12 | Kitano et al, 2009 |
| | Siamang | Symphalangus syndactylus | B | 2 | Socha & Moor-Jankowski, 1979 | B | 6 | Kitano et al, 2009 |

| Species | n | S | Syn | Nsyn |
|---|---|---|---|---|
| *Homo sapiens* | 179 | 15 | 6 | 9 |
| *Pan paniscus* | 11 | 2 | 0 | 2 |
| *Pan troglodytes* | 35 | 7 | 3 | 4 |
| *Gorilla gorilla* | 31 | 2 | 1 | 1 |
| *Pongo pygmaeus* | 12 | 5 | 2 | 3 |
| *Symphalangus syndactylus* | 6 | 2 | 0 | 2 |
| *Hylobates agilis* | 5 | 13 | 5 | 8 |
| *Hylobates lar* | 12 | 11 | 4 | 7 |
| *Papio anubis* | 31 | 8 | 4 | 4 |
| *Macaca fascicularis* | 7 | 17 | 10 | 7 |
| *Macaca mulatta* | 6 | 13 | 9 | 4 |
| *Chlorocebus aethiops* | 4 | 1 | 0 | 1 |
| *Colobus angolensis* | 3 | 14 | 6 | 8 |
| *Colobus polykomos* | 1 | 13 | 7 | 6 |
| *Colobus guerenza* | 1 | 13 | 7 | 6 |
| *Callithrix jacchus* | 4 | 2 | 2 | 0 |
| *Alouatta caraya* | 3 | 0 | 0 | 0 |
| **Total across the dataset** | **351** | **68** | **36** | **32** |

| Species | Forward Primer | Reverse Primer | Annealing Temperature |
|---|---|---|---|
| Hominoids | TCTGCTGCTCTAAGCCTTCC | CCCGTTCTGCTAAAACCAAG | 58.4 |
| Marmosets, Vervets monkeys, Colobus monkeys | TGGTGCATCTGCTSCTCTGAG | CCRTTCTKCTAAAACCAAGGG | 58.0 |
| Howler Monkeys | CCACCGGGTCCACTACTACA | TTGGTGGGTTTGTGGCGCAGT | 60.0 |

| Nucleotide position | 557 | 559 | 567 | 568 | 569 | 574 | 579 | 603 | 613 | 621 | 627 | 633 | 638 | 659 | 681 | 689 | 722 | 727 | 745 | 759 | 780 | 802 | 803 | 811 | 816 | 819 | 857 | 873 | 891 | 893 | 903 | 913 | 926 | 947 |
|---|---|---|---|---|---|---|---|---|---|---|---|---|---|---|---|---|---|---|---|---|---|---|---|---|---|---|---|---|---|---|---|---|---|---|
| Chimp_KOMINATO_01 | A | C | C | A | G | A | G | A | G | A | G | A | C | C | G | C | G | C | G | C | C | T | G | C | G | C | T | C | A | C | G | A | G | G |
| Chimp_KOMINATO_02 | A | C | C | A | G | A | G | A | G | A | G | A | C | C | G | C | G | C | G | C | C | T | G | C | G | C | C | C | A | C | G | A | G | G |
| Chimp_KERMARREC_A1 | A | C | C | A | G | A | G | A | G | A | G | A | C | C | G | C | G | C | G | C | C | T | G | C | G | C | C | C | A | C | G | A | G | G |
| Chimp_KERMARREC_A2 | A | C | C | A | G | A | G | A | G | A | G | A | C | C | G | C | G | C | G | C | C | T | G | C | G | C | T | C | A | C | G | A | G | G |
| Chimp_KERMARREC_Ox | A | C | C | A | G | A | G | A | G | A | G | A | C | C | G | C | G | C | G | C | C | T | G | C | G | C | C | C | G | C | G | A | G | G |
| Chimp_KERMARREC_Odel | A | C | C | A | G | A | G | A | G | A | G | C | C | G | G | C | G | C | G | C | C | T | G | C | G | C | T | C | A | C | G | A | G | G |
| Chimp_SUMIYAMA_01 | A | C | C | A | G | A | G | A | G | A | G | A | C | C | G | C | G | C | G | C | C | T | G | C | G | C | T | C | A | C | G | A | G | G |
| Chimp_SUMIYAMA_02 | A | C | C | A | G | A | G | A | G | A | G | A | C | C | G | C | G | C | G | C | C | T | G | C | G | C | C | C | A | C | G | A | G | G |
| Chimp_SUMIYAMA_03 | A | C | C | A | G | A | G | A | G | A | G | A | C | C | G | C | G | C | G | C | C | T | G | C | G | C | C | C | G | C | G | A | G | G |
| Chimp_SUMIYAMA_04 | A | C | C | A | G | A | G | A | G | A | G | A | C | C | G | C | G | C | G | C | C | T | G | C | G | C | T | C | G | C | G | A | G | G |
| Chimp_SUMIYAMA_05 | A | C | C | A | G | A | G | A | G | A | G | C | C | G | G | C | G | C | G | C | C | T | G | C | G | C | T | C | A | C | G | A | G | G |
| Chimp_ThisStudy_01 | A | C | C | A | G | A | G | A | G | A | G | A | C | C | G | C | G | C | G | C | C | T | G | C | G | C | T | C | A | C | G | A | G | G |
| Chimp_ThisStudy_02 | A | C | C | A | G | A | G | A | G | A | G | A | C | C | G | C | G | C | G | C | C | T | G | C | G | C | C | T | A | C | G | A | G | G |
| Chimp_ThisStudy_03 | A | C | C | A | G | A | G | A | G | A | G | A | C | C | G | C | G | C | G | C | G | T | G | C | G | C | C | C | G | C | G | A | G | G |
| Chimp_ThisStudy_04 | A | C | C | A | G | A | G | A | G | A | G | A | C | C | G | C | G | C | G | C | C | T | G | C | G | C | C | C | G | C | G | A | G | G |
| Chimp_ThisStudy_05 | A | C | C | A | G | A | G | A | G | A | G | A | C | C | G | C | C | G | C | G | C | C | T | G | C | G | C | C | C | A | C | G | A | G | G |
| Chimp_ThisStudy_06 | A | C | C | A | G | A | G | A | G | A | G | A | C | C | G | T | G | C | C | T | G | C | G | C | C | C | A | C | G | A | G | G |
| Chimp_ThisStudy_07 | A | C | C | A | G | A | G | A | G | A | G | C | C | G | G | C | G | C | G | C | C | T | G | C | G | C | T | C | A | C | G | A | G | G |
| Chimp_ThisStudy_08 | A | C | C | A | G | A | G | A | G | A | G | C | C | G | G | C | G | C | G | C | C | T | G | C | G | C | C | C | A | C | G | A | G | G |
| Chimp_MARTINKO_01 | A | C | C | G | G | G | G | A | G | A | G | A | C | G | G | C | G | C | C | C | G | C | G | C | T | C | A | C | G | A | G | G |
| Chimp_MARTINKO_02 | A | C | T | G | G | G | G | A | G | A | G | A | C | C | G | C | G | C | G | C | C | T | G | C | G | C | C | C | A | C | G | A | G | G |
| Chimp_MARTINKO_03 | A | C | G | C | G | C | A | G | A | G | A | C | C | G | C | G | C | G | C | C | T | G | C | G | C | C | C | A | C | G | A | G | G |
| Gorilla_KOMINATO | A | C | C | G | G | G | G | A | G | A | G | A | C | G | G | C | G | C | C | C | G | C | G | C | C | C | A | A | C | G | G | G |
| Gorilla_ThisStudy_01 | A | C | C | G | G | G | G | A | G | A | G | A | C | G | G | C | G | C | C | C | A | C | G | C | C | C | A | A | C | G | G | G |
| Gorilla_ThisStudy_02 | A | C | C | G | G | G | G | A | G | A | G | A | C | G | G | C | G | C | C | C | G | C | G | C | C | C | A | A | C | G | G | G |
| Gorilla_ThisStudy_03 | A | C | C | G | G | G | G | A | G | A | G | A | C | G | G | C | G | C | C | G | T | G | C | G | C | C | C | A | A | C | G | G | G |
| Gorilla_MARTINKO_01 | A | C | C | G | G | G | G | A | G | A | G | A | C | G | G | C | G | C | C | G | C | G | C | C | C | A | A | C | G | G | G |
| Gorilla_MARTINKO_02 | A | C | C | G | G | G | G | A | G | A | G | A | C | G | G | C | G | C | C | C | G | C | G | C | C | C | A | A | C | G | A | G |
| Gorilla_MARTINKO_03_04 | A | C | C | G | G | G | G | A | G | A | G | A | C | G | C | C | C | G | C | G | C | C | C | A | A | C | G | G | G |
| Gorilla_MARTINKO_05 | A | C | C | G | G | G | G | A | G | A | C | G | G | C | T | C | C | C | G | C | G | C | C | C | A | A | C | G | G | G |
| Orang_ThisStudy_01 | A | C | C | G | G | G | G | A | G | G | A | C | G | A | C | G | T | C | C | G | C | C | T | C | C | A | C | G | A | G | G |
| Orang_ThisStudy_02 | A | C | C | G | G | G | G | A | G | G | A | C | G | A | C | G | T | C | C | G | C | C | T | C | C | A | C | G | A | G | A |
| Orang_ThisStudy_03 | A | C | C | G | G | G | G | A | G | G | A | C | G | A | C | G | T | C | C | G | C | C | T | C | C | A | A | C | A | G | G |
| Orang_ThisStudy_04 | A | T | C | G | G | G | G | A | G | G | A | C | G | A | C | G | T | C | C | A | C | C | T | C | C | A | A | C | A | G | G |
| Orang_ThisStudy_05 | A | C | C | G | G | G | G | A | G | G | A | C | G | A | C | G | T | C | C | A | C | C | T | C | C | A | A | C | A | G | G |
| Orang_KOMINATO_01 | A | C | C | G | G | G | G | A | G | G | A | C | G | A | C | G | T | C | C | G | C | C | T | C | C | A | C | G | A | G | A |
| Orang_KOMINATO_02 | A | C | C | G | G | G | G | A | G | G | G | A | C | G | A | C | G | T | C | C | G | C | C | T | C | C | A | C | G | A | G | G |
| Orang_MARTINKO | T | C | C | G | G | G | T | T | G | T | A | G | G | A | C | G | T | C | C | G | C | C | C | C | C | A | C | G | G | G | G |